\documentclass[reprint,twocolumn]{revtex4}
\usepackage{amsfonts}
\usepackage{amsmath}
\usepackage{amssymb}
\usepackage{charter}
\usepackage{graphicx}
\usepackage{float}
\usepackage{amsmath}
\usepackage{amssymb}
\usepackage{charter}
\usepackage{graphicx}
\usepackage{subfigure}
\usepackage{graphicx}
\usepackage{epstopdf}
\usepackage{color}
\usepackage{tocvsec2}
\usepackage{enumerate}
\usepackage{graphicx}
\usepackage{dcolumn}
\usepackage{bm}

\begin{document}

\title{Phase estimation of Mach-Zehnder interferometer via Laguerre excitation squeezed
state }
\author{Zekun Zhao$^{1}$}
\author{Huan Zhang$^{2}$}
\author{Yibing Huang$^{1}$}
\thanks{huangyb@aliyun.com}
\author{Liyun Hu$^{1}$}
\thanks{hlyun@jxnu.edu.cn}
\affiliation{$^{{\small 1}}$\textit{Center for Quantum Science and Technology, Jiangxi
Normal University, Nanchang 330022, China}\\
$^{{\small 2}}$\textit{School of Physics, Sun Yat-sen University, Guangzhou
510275, China}}

\begin{abstract}
Quantum metrology has an important role in the fields of quantum optics and
quantum information processing. Here we introduce a kind of non-Gaussian
state, Laguerre excitation squeezed state as input of traditional
Mach-Zehnder interferometer to examine phase estimation in realistic case.
We consider the effects of both internal and external losses on phase
estimation by using quantum Fisher information and parity detection. It is
shown that the external loss presents a bigger effect than the internal one.
The phase sensitivity and the quantum Fisher information can be improved by
increasing the photon number and even surpass the ideal phase sensitivity by
two-mode squeezed vacuum in a certain region of phase shift for realistic
case.

\end{abstract}

\maketitle

\section{Introduction}

Optical quantum metrology is one of the most important branches in the field
of quantum science, which plays a key role for the advanced
development of science and technology application. It is characteristics of
using quantum systems or quantum mechanical properties, such as
entanglement, squeezing or nonclassical property, to achieve high precision
measurements of physical parameters, by minimizing the measurement
uncertainty. It is shown that the precision of measurement can break through
the standard quantum limit (SQL) due to the quantum effects. Based on this
interesting point, the researchers focus their attention on the improvement
of measurement precision by using quantum properties.

To realize this purpose above, the Mach-Zehnder interferometer (MZI) is
widely used in various tasks of quantum measurement \cite{1,2,3,4}.
Generally, the measurement process can be divided into three parts, i.e.,
the preparation of input states, the interaction between the input state and
the considered system, and the detection on the output state \cite{5,6}.
Thus, it is natural to examine these three parts separately or collectively
for enhancing the measurement precision. For instance, when injecting
separately the coherent state and the squeezed state into two input ports of
MZI \cite{7}, the phase precision can beat the SQL of $1/\sqrt{\overline{N}}$%
, with $\overline{N}$ being the average photon-number of the input state.
After that, many different quantum states have been proposed as the input
states of MZI to achieve better performance. Among them, the NOON state \cite%
{8}, twin Fock state \cite{9}, and the two-mode squeezed vacuum state (TMSV)
\cite{10} \textit{et al.} can achieve or even exceed the Heisenberg limit
(HL) $1/\overline{N}$ \cite{11,12,13}, which have been verified by many
experiments\ \cite{14,15}. However, on one hand, it is difficult to prepare
a high average photon-number of quantum states of light \cite{16}. On the
other hand, the precision will be quickly destroyed due to the inevitable
interaction between the systems and the environments \cite{16,17,18,19,20,21}%
. For example, for the TMSV, the experimentally available squeezing
parameter is approximately $1.15$ corresponding to a small average
photon-number about $2\sinh ^{2}r\approx 4$ \cite{16}. In addition, the
phase sensitivity is unstable relative to the phase shift. That is to say,
the phase sensitivity will deteriorate rapidly when deviating from the
optimal phase shift \cite{22}. Thus, it is still a challenging task how to
further improve the measurement precision and the ability against the
decoherence.

Actually, the high nonclassical property including entanglement play an
important role in various quantum information tasks, including quantum key
distribution \cite{23}, quantum teleportation \cite{24}, and quantum
metrology \cite{7,8,9,10,25,26,27,28,29,30,31,32,33,34,35,36}. Thus,
preparing a kind of high nonclassical property state as inputs is an
effective method to improve the measurement precision. For example, mixing
photon-added/subtracted squeezed vacuum and coherent state as inputs, it
is found that the phase sensitivity can be improved \cite{25,32,33,34}.
Using photon-added/subtracted TMSV input the MZI can improve the precision
of phase estimation \cite{26,27}. Recently, by employing multi-photon
catalysis (MC) operating on the TMSV (MC-TMSV) as inputs of MZI \cite{35},
Zhang \emph{et al}. studied the phase measurement including the case of
photon losses. It is shown that the influences of photon losses before
parity detection (external dissipation) on phase measurement accuracy is
more serious than that after phase shifter (internal dissipation), but these
effects can be suppressed by increasing the number of catalytic photons. In
addition, the photon-number conversing operation is also used to improve
phase estimation \cite{36}.

These above research works indicate that non-Gaussian operation is an
effective way to improve the measurement precision. Inspired by this, we
introduce a kind of non-Gaussian operation, i.e., Laguerre polynomial
excitation operating on the TMSV as inputs of MZI, to improve the phase
sensitivity. In fact, Laguerre polynomial excitation can achieve high
nonclassicality and be theoretically realized \cite{37,38}. We shall
investigate the phase sensitivity with parity detection and the quantum
Fisher information (QFI) in both ideal and realistic cases, by deriving an
equivalent operator with the aid of the Weyl ordering invariance under
similarity transformations. It is found that the phase sensitivity and the
QFI can be improved whose effects become more obvious as the excited order.

This paper is organized as follows. In Sec. II, we first introduce the
Laguerre polynomial excitation squeezed state. Then, we examine the QFI and
the phase sensitivity with parity measurement in ideal case, when
considering Laguerre polynomial excitation squeezed state as inputs. In Sec.
III, we further consider the effects of photon losses on the phase
sensitivity including external and internal dissipations. In Sec. IV, we
investigate the influence of photon losses on the QFI. The main results are
summarized in the last section.

\section{Phase estimation when the Laguerre polynomial excitation squeezed
state as inputs of MZI in ideal case}

\subsection{Laguerre polynomial excitation squeezed state as Two-mode
squeezed twin-Fock state}

Actually, the Laguerre polynomial excitation squeezed state $\left\vert
\text{Lagu}\right\rangle $ can be generated by applying two-mode squeezing
operator on twin-Fock state $\left\vert n,n\right\rangle $, i.e.,
\begin{equation}
\left\vert \text{Lagu}\right\rangle =S\left( r\right) \left\vert
n,n\right\rangle ,  \label{1}
\end{equation}%
where $S\left( r\right) =\exp \left\{ r(a^{\dagger }b^{\dagger }-ab)\right\}
$ is the two-mode squeezing operator and $\left\vert n,n\right\rangle
=\left\vert n\right\rangle _{a}\otimes \left\vert n\right\rangle _{b}$ is
twin-Fock state. Using the coherent state representation of Fock state,
i.e.,
\begin{equation}
\left\vert n\right\rangle _{a}=\left. \frac{\partial ^{n}}{\sqrt{n!}\partial
\tau ^{n}}\left\Vert \tau \right\rangle \right\vert _{\tau =0},\left\Vert
\tau \right\rangle =e^{\tau a^{\dag }}\left\vert 0\right\rangle _{a},
\label{2}
\end{equation}%
and the transform relations
\begin{eqnarray}
S\left( r\right) aS^{\dag }\left( r\right)  &=&a\cosh r-b^{\dag }\sinh r,
\notag \\
S\left( r\right) bS^{\dag }\left( r\right)  &=&b\cosh r-a^{\dag }\sinh r,
\label{3}
\end{eqnarray}%
Eq. (\ref{1}) can be rewritten as the following form
\begin{equation}
\left\vert \text{Lagu}\right\rangle =\left( -\tanh r\right) ^{n}L_{n}\left(
ua^{\dag }b^{\dag }\right) S\left\vert 00\right\rangle ,  \label{4}
\end{equation}%
where we have used $u=2/\sinh 2r$, $S\left\vert 00\right\rangle =$sech$r\exp
\left\{ a^{\dag }b^{\dag }\tanh r\right\} \left\vert 00\right\rangle $ and
the formula $e^{A+B}=e^{A}e^{B}e^{-1/2[A,B]},$ which is valid for $%
[A,[A,B]]=[B,[A,B]]=0,$ as well as $e^{\lambda a}a^{\dag }e^{-\lambda
a}=a^{\dag }+\lambda ,$ and
\begin{equation}
L_{n}\left( xy\right) =\frac{\left( -1\right) ^{n}}{n!}\frac{\partial ^{2n}}{%
\partial \tau ^{n}\partial t^{n}}e^{-\tau t+\tau x+ty}|_{\tau =t=0},
\label{5}
\end{equation}%
with $L_{n}\left( xy\right) $ being Laguerre polynomials. From Eq. (\ref{4})
it is clear that Laguerre polynomial excitation squeezed state is just the
two-mode squeezed Fock state \cite{39}. It is interesting that the twin-Fock
states with 6 photons can be achieved experimentally \cite{14,40}. Thus, the
Laguerre polynomial excitation squeezed state can be successfully realized.

Using Eq. (\ref{1}) and Eq. (\ref{3}) it is ready to have the total average
photon number, i.e.,
\begin{eqnarray}
\bar{N} &=&\left \langle \text{Lagu}\right \vert \left( a^{\dag }a+b^{\dag
}b\right) \left \vert \text{Lagu}\right \rangle  \notag \\
&=&2n\cosh 2r+2\sinh ^{2}r.  \label{6}
\end{eqnarray}%
It is clear that the total average photon number of input state increases
with $r$ and $n$.

\subsection{Laguerre polynomial excitation squeezed state as input of MZI
and parity detection}

In order to establish the basis of studying the phase estimation via
Laguerre polynomial excitation squeezed state in the non-ideal case, here we
consider the Laguerre polynomial excitation squeezed state as input of MZI
for discussing the effect of this non-Gaussian state on the precision of
measurement in the ideal case. As shown in Fig. 1, the traditional MZI
consist of two symmetrical beam splitters (BSs) (denoted as BS1 and BS2),
two input ports (mode $a$ and $b$) and two completely reflecting mirrors as
well as two-phase shifters. Here we should note that the two BSs are
conjugated to each other.

For this ideal MZI in Fig. 1, according to Ref. \cite{41}, the effect is
equivalent to a BS operator, i.e.,
\begin{equation}
U_{MZI}=e^{i\pi J_{1}/2}e^{-i\varphi J_{3}}e^{-i\pi J_{1}/2}=e^{-i\varphi
J_{2}},  \label{7}
\end{equation}%
where $J_{1},J_{2},J_{3}$ are Bosonic operators, defined as%
\begin{eqnarray}
J_{1} &=&\frac{1}{2}\left( a^{\dag }b+ab^{\dag }\right) ,  \notag \\
J_{2} &=&\frac{1}{2i}\left( a^{\dag }b-ab^{\dag }\right) ,  \notag \\
J_{3} &=&\frac{1}{2}\left( a^{\dag }a-b^{\dag }b\right) .  \label{8}
\end{eqnarray}

\begin{figure}[tph]
\label{Fig1} \centering \includegraphics[width=0.83\columnwidth]{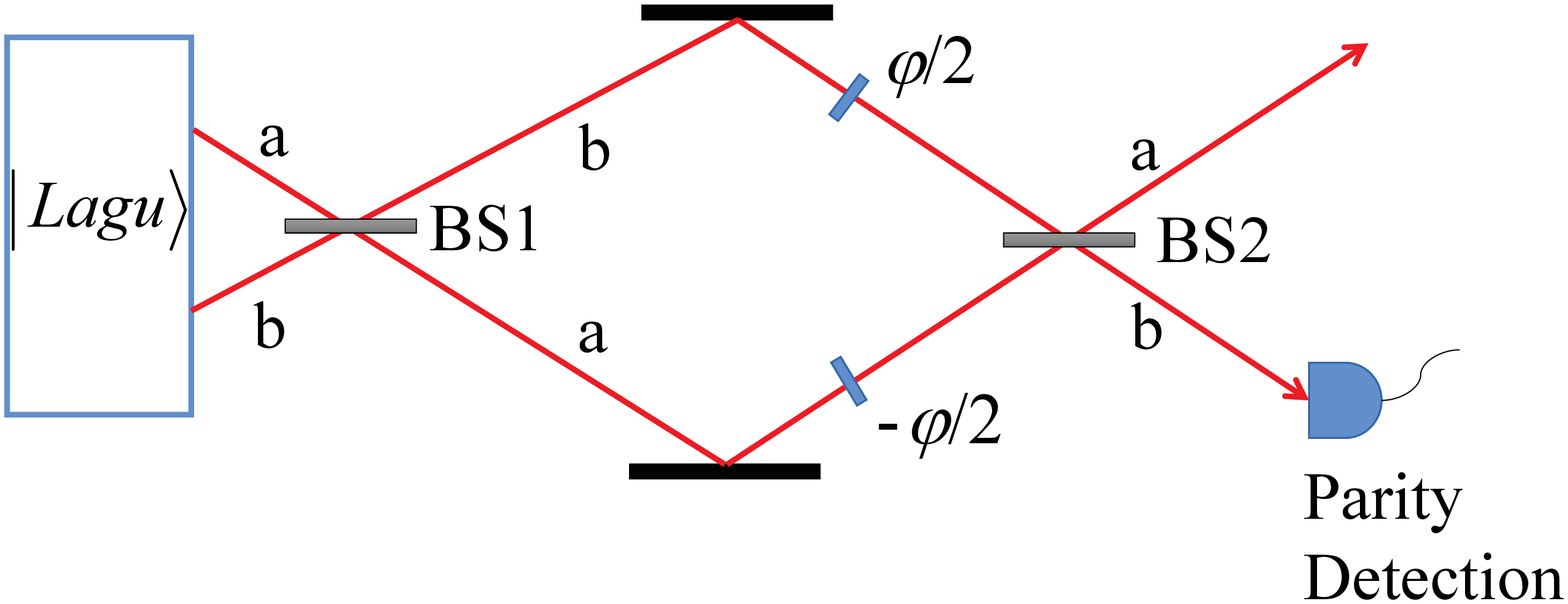}
\caption{{}Schematic diagram of a balanced MZI for the parity detection of
the phase shift when the Laguerre excitation squeezed state is injected into
the first beam splitter. }
\end{figure}

\subsubsection{The quantum Fisher information}

Here we examine the QFI when inputting $\left\vert \text{Lagu}\right\rangle $
into the MZI in ideal case. For the model shown in Fig. 1, the QFI $F_{Q}$
describes the amount of information containing phase parameters carried by
light after it passes through the phase shifter. The quantum Cram\'{e}r-Rao
bound (QCRB) gives the highest theoretical measurement accuracy of phase
shifts, which is expressed by the QFI \cite{6}, i.e.,

\begin{equation}
\Delta \varphi _{QCRB}=\frac{1}{\sqrt{F_{Q}}}.  \label{9}
\end{equation}%
For the pure state as the input state $\left \vert in\right \rangle $ of
MZI, the QFI $F_{Q}$ can be calculated by

\begin{equation}
F_{Q}=4\left[ \left \langle \psi ^{\prime }\left( \varphi \right) |\psi
^{\prime }\left( \varphi \right) \right \rangle -\left \vert \left \langle
\psi ^{\prime }\left( \varphi \right) |\psi \left( \varphi \right) \right
\rangle \right \vert ^{2}\right] ,  \label{10}
\end{equation}%
where $\left \vert \psi \left( \varphi \right) \right \rangle =e^{-i\varphi
J_{3}}e^{-i\pi J_{1}/2}\left \vert in\right \rangle $ is the quantum state
after the evolution of the first BS and phase shifter, $\left \vert \psi
^{\prime }\left( \varphi \right) \right \rangle =\partial \left \vert \psi
^{\prime }\left( \varphi \right) \right \rangle /\partial \varphi $.
Therefore, it can be known that in the case of $\left \vert \text{Lagu}%
\right \rangle $ as the input state of MZI, the expression of the QFI can be
derived\ as

\begin{equation}
F_{Q}=\left[ 2+3\sinh ^{2}\left( 2r\right) \right] n\left( n+1\right) +\sinh
^{2}\left( 2r\right) .  \label{11}
\end{equation}

\subsubsection{The phase sensitivity with parity detection}

Through this paper, we shall take parity detection\ as measurement method.
Here, we consider the parity detection at output mode $b$. Actually, the
photon-number parity operator is given by%
\begin{equation}
\Pi _{b}=\left( -1\right) ^{b^{\dag }b}=e^{i\pi b^{\dag }b},  \label{12}
\end{equation}%
whose normal ordering form is
\begin{equation}
\Pi _{b}=\colon \exp \left\{ -2b^{\dag }b\right\} \colon ,  \label{13}
\end{equation}%
where $\colon \cdot \colon $ is the symbol of the normal ordering. Thus
using the formula converting operator $\hat{O}$ from normal ordering to its
Weyl ordering form, i.e.,
\begin{equation}
\hat{O}=2%
\begin{array}{c}
\colon  \\
\colon
\end{array}%
\int \frac{d^{2}\alpha }{\pi }\left\langle -\alpha \right\vert \hat{O}%
\left\vert \alpha \right\rangle e^{2\left( \alpha ^{\ast }b-b^{\dag }\alpha
+b^{\dag }b\right) }%
\begin{array}{c}
\colon  \\
\colon
\end{array}%
,  \label{14}
\end{equation}%
where $\left\vert \alpha \right\rangle $ is the coherent state, the Weyl
ordering form of parity operator $\Pi _{b}$ can be derived as%
\begin{equation}
\Pi _{b}=\frac{\pi }{2}%
\begin{array}{c}
\colon  \\
\colon
\end{array}%
\delta \left( b\right) \delta \left( b^{\dag }\right)
\begin{array}{c}
\colon  \\
\colon
\end{array}%
,  \label{15}
\end{equation}%
where$%
\begin{array}{c}
\colon  \\
\colon
\end{array}%
\cdots
\begin{array}{c}
\colon  \\
\colon
\end{array}%
$ is the symbol of the Weyl ordering and $\delta \left( \cdot \right) $ is
the delta function \cite{42,43}.

Noticing the Weyl ordering invariance under similarity transformations \cite%
{44,45}, i.e.,
\begin{equation}
U_{MZI}^{\dag }%
\begin{array}{c}
\colon  \\
\colon
\end{array}%
\cdots
\begin{array}{c}
\colon  \\
\colon
\end{array}%
U_{MZI}=%
\begin{array}{c}
\colon  \\
\colon
\end{array}%
U_{MZI}^{\dag }\cdots U_{MZI}%
\begin{array}{c}
\colon  \\
\colon
\end{array}%
,  \label{16}
\end{equation}%
and the transformation relations%
\begin{eqnarray}
e^{i\varphi J_{2}}ae^{-i\varphi J_{2}} &=&a\cos \frac{\varphi }{2}-b\sin
\frac{\varphi }{2},  \notag \\
e^{i\varphi J_{2}}be^{-i\varphi J_{2}} &=&b\cos \frac{\varphi }{2}+a\sin
\frac{\varphi }{2},  \label{17}
\end{eqnarray}%
then the parity operator under the unitary transformation is changed to be%
\begin{eqnarray}
\Pi _{b}\left. \rightarrow \right. \Pi _{MZI} &\equiv &U_{MZI}^{\dag }\Pi
_{b}U_{MZI}  \notag \\
&=&\frac{\pi }{2}%
\begin{array}{c}
\colon  \\
\colon
\end{array}%
\delta \left( b\cos \frac{\varphi }{2}+a\sin \frac{\varphi }{2}\right)
\notag \\
&&\times \delta \left( b^{\dag }\cos \frac{\varphi }{2}+a^{\dag }\sin \frac{%
\varphi }{2}\right)
\begin{array}{c}
\colon  \\
\colon
\end{array}%
,  \label{18}
\end{eqnarray}%
which is just the Weyl ordering form of the parity operator $\Pi _{b}$ under
the unitary transformation $U_{MZI}$.

For a Weyl ordering operator, say $%
\begin{array}{c}
\colon  \\
\colon
\end{array}%
f\left( a,a^{\dag },b,b^{\dag }\right)
\begin{array}{c}
\colon  \\
\colon
\end{array}%
$, its classical correspondence can be obtained by replacing $a,a^{\dag
},b,b^{\dag }$ with complex parameters $\alpha ,\alpha ^{\ast },\beta ,\beta
^{\ast },$ respectively, i.e., $%
\begin{array}{c}
\colon  \\
\colon
\end{array}%
f\left( a,a^{\dag },b,b^{\dag }\right)
\begin{array}{c}
\colon  \\
\colon
\end{array}%
\rightarrow f\left( \alpha ,\alpha ^{\ast },\beta ,\beta ^{\ast }\right) $.
Further using the relation between classical correspondence and Wigner
operator \cite{45}, i.e.,
\begin{eqnarray}
&&%
\begin{array}{c}
\colon  \\
\colon
\end{array}%
f\left( a,a^{\dag },b,b^{\dag }\right)
\begin{array}{c}
\colon  \\
\colon
\end{array}
\notag \\
&=&4\int d^{2}\alpha d^{2}\beta f\left( \alpha ,\alpha ^{\ast },\beta ,\beta
^{\ast }\right) \Delta _{a}\left( \alpha \right) \Delta _{b}\left( \beta
\right) ,  \label{19}
\end{eqnarray}%
where $\Delta _{a/b}\left( \alpha /\beta \right) $ is the Wigner operators
whose normal ordering form is given by \cite{46,47}%
\begin{eqnarray}
\Delta _{a}\left( \alpha \right)  &=&\frac{1}{\pi }\colon \exp \left[
-2\left( a-\alpha \right) \left( a^{\dag }-\alpha ^{\ast }\right) \right]
\colon ,  \notag \\
\Delta _{b}\left( \beta \right)  &=&\frac{1}{\pi }\colon \exp \left[
-2\left( b-\beta \right) \left( b^{\dag }-\beta ^{\ast }\right) \right]
\colon ,  \label{20}
\end{eqnarray}%
and\ using the integration within an ordered product (IWOP) technique \cite%
{47,48} as well as the following integral formula \cite{49}%
\begin{equation}
\int \frac{d^{2}z}{\pi }e^{\zeta \left\vert z\right\vert ^{2}+\xi z+\eta
z^{\ast }+fz^{2}+gz^{\ast 2}}=\frac{e^{\frac{-\zeta \xi \eta +\xi ^{2}g+\eta
^{2}f}{\zeta ^{2}-4fg}}}{\sqrt{\zeta ^{2}-4fg}},  \label{21}
\end{equation}%
the normal ordering of $\Pi _{MZI}$ can be obtained, i.e.,

\begin{eqnarray}
\Pi _{MZI} &=&\colon \exp \left[ \left( -\sin \varphi -1\right) a^{\dagger
}a+\left( \sin \varphi -1\right) b^{\dagger }b\right]   \notag \\
&&\times \exp \left[ -\left( b^{\dag }a+a^{\dag }b\right) \cos \varphi %
\right] \colon .  \label{22}
\end{eqnarray}%
According to Ref. \cite{10}, we has made a shift transformation $\varphi
\longrightarrow \varphi +\pi /2$ in Eq. (\ref{22}). When the state $%
\left\vert \text{Lagu}\right\rangle $ as input of MZI, the expectation value
of parity operator in the output state can be expressed as $\left\langle \Pi
_{0}\right\rangle =\left\langle \text{Lagu}\right\vert U_{MZI}^{\dag }\Pi
_{b}U_{MZI}\left\vert \text{Lagu}\right\rangle =\left\langle \text{Lagu}%
\right\vert \Pi _{MZI}\left\vert \text{Lagu}\right\rangle $, where $\Pi
_{MZI}=U_{MZI}^{\dag }\Pi _{b}U_{MZI}$ whose normal ordering form is given
in Eq. (\ref{22}). Inserting the completeness relation of coherent state and
using Eq. (\ref{21}), $\left\langle \Pi _{0}\right\rangle $ can be
calculated as
\begin{eqnarray}
\left\langle \Pi _{0}\right\rangle  &=&A_{0}\hat{D}_{n}\{\exp \left[
(x^{2}+t^{2}-y^{2}-\tau ^{2})A_{1}\right]   \notag \\
&&\times \exp \left[ (xy+t\tau )A_{2}\right]   \notag \\
&&\times \exp \left[ (y\tau -xt)A_{3}\right]   \notag \\
&&\times \exp \left[ (x\tau +yt)A_{4}\right] \},  \label{23}
\end{eqnarray}%
where $\hat{D}_{n}\left\{ \cdot \right\} =\frac{\partial ^{4n}}{\left(
n!\right) ^{2}\partial x^{n}\partial y^{n}\partial t^{n}\partial \tau ^{n}}%
\left\{ \cdot \right\} |_{x=y=t=\tau =0},$ and

\begin{eqnarray}
A_{0} &=&\frac{\text{sech}^{2}r}{\sqrt{\omega _{0}}},  \notag \\
A_{1} &=&\frac{\sin \left( 2\varphi \right) \tanh r}{2\omega _{0}\cosh ^{2}r}%
,  \notag \\
A_{2} &=&\frac{\left( \cos \left( 2\varphi \right) -1\right) \left( \tanh
r+\tanh ^{3}r\right) }{\omega _{0}},  \notag \\
A_{3} &=&\frac{\sin \varphi \cosh \left( 2r\right) \text{sech}^{4}r}{\omega
_{0}},  \notag \\
A_{4} &=&\frac{-\cos \varphi \text{sech}^{4}r}{\omega _{0}},  \label{24}
\end{eqnarray}%
as well as $\omega _{0}=1-2\tanh ^{2}r\cos \left( 2\varphi \right) +\tanh
^{4}r$. Thus, using Eq. (\ref{23}) we can get the phase sensitivity $%
\triangle \varphi _{0}$ via error propagation formula, i.e.,

\begin{equation}
\triangle \varphi _{0}=\frac{\triangle \Pi _{0}}{\left\vert \partial
\left\langle \Pi _{0}\right\rangle /\partial \varphi \right\vert },
\label{25}
\end{equation}%
where $\triangle \Pi _{0}=\sqrt{1-\left\langle \Pi _{0}\right\rangle ^{2}}$.
From the value of $\triangle \varphi _{0}$, in principle, we can know the
phase measurement accuracy of Lagurre polynomial excitation squeezed state as input of
MZI.

In particular, when $n=0$ corresponding to the TMSV as input of MZI, the
phase sensitivity with parity detection is given by
\begin{equation}
\triangle \varphi _{TMSV}=\frac{\omega _{0}\cosh ^{2}r}{2\tanh r\cos \varphi
},  \label{26}
\end{equation}%
as expected \cite{10}.

\section{Effects of photon losses on phase sensitivity}

In the process of quantum precision measurement, photon losses is
inevitable. It is of great practical significance to study the influence of
photon losses on phase sensitivity. In this section, we consider the phase
sensitivity with Laguerre polynomial excitation squeezed state as input of MZI in
photon losses case. Here, we only examine that the photon losses occurs
either before parity detection in MZI (outside the interferometer) or
between the phase shift and the second BS (inside the interferometer), shown
in Fig. 2.
\begin{figure}[tbh]
\label{Fig2} \centering
\subfigure{
\begin{minipage}[b]{0.5\textwidth}
\includegraphics[width=0.83\textwidth]{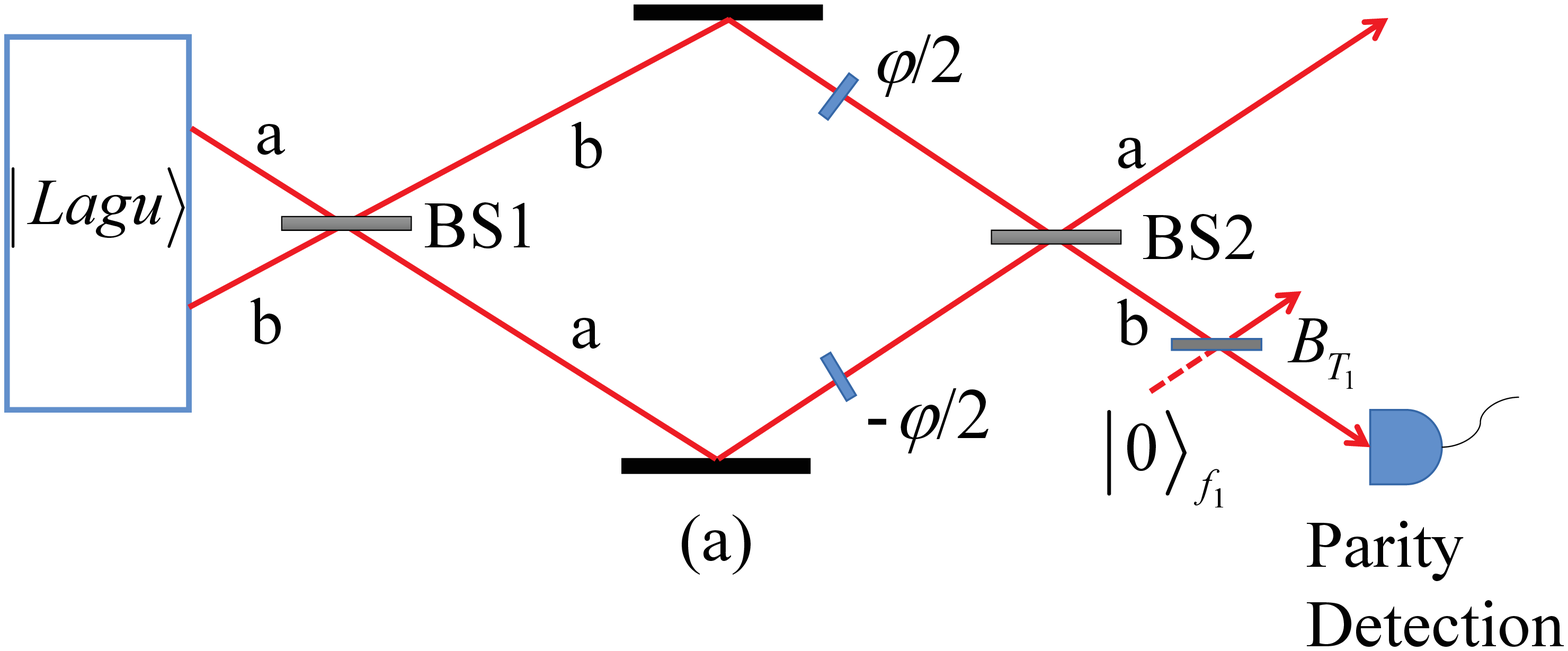}\\
\includegraphics[width=0.83\textwidth]{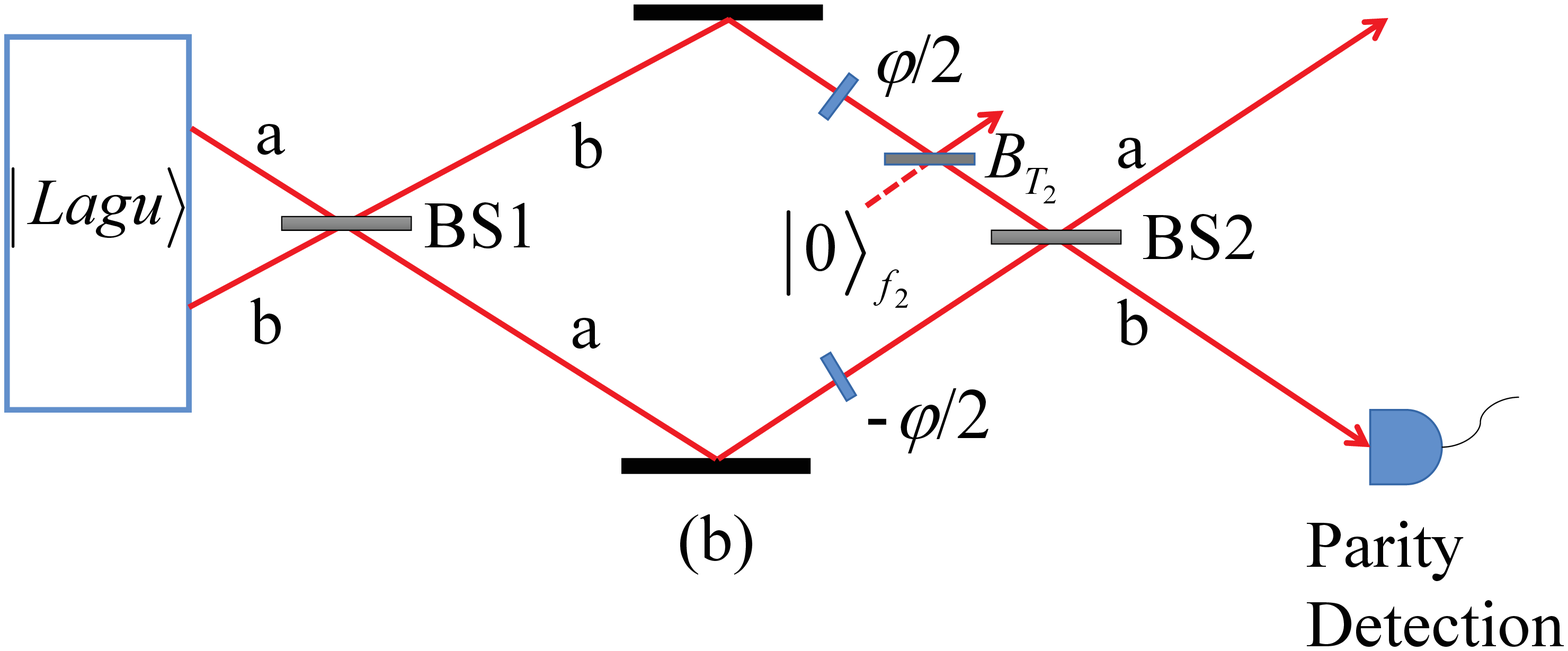}
\end{minipage}}
\caption{Schematic diagram of the parity detection in the presence of photon
losses. (a) External dissipation: photon losses occur between the\ parity
detection and the BS2. (b) Internal dissipation: photon losses occur between
the phase shifter and the BS2.}
\end{figure}

\subsection{Effects of photon losses before parity detection (external
dissipation)}

First, we focus on the case with photon losses before parity detection as
shown in Fig. 2 (a) and investigate the effects of photon losses on \ the
phase sensitivity. It will be convenient to redefine an equivalent parity
operator including photon losses, which is different from the ideal case
where parity operator is $\Pi _{b}=\left( -1\right) ^{b^{\dag }b}$. For this
purpose, we use an optical BS $B_{T_{1}}$ to simulate the photon losses at
the probe end, shown in Fig. 2(a). The corresponding transform relation by $%
B_{T_{1}}$ is given by%
\begin{equation}
B_{T_{1}}^{\dagger }\left(
\begin{array}{c}
b \\
f_{1}%
\end{array}%
\right) B_{T_{1}}=\left(
\begin{array}{cc}
\sqrt{T_{1}} & \sqrt{1-T_{1}} \\
-\sqrt{1-T_{1}} & \sqrt{T_{1}}%
\end{array}%
\right) \left(
\begin{array}{c}
b \\
f_{1}%
\end{array}%
\right) ,  \label{27}
\end{equation}%
where $f_{1}$ ($f_{1}^{\dagger }$) are photon annihilation (creation)
operators corresponding to the dissipative mode $f_{1}$ of $B_{T_{1}}$ and $%
T_{1}$ is the transmissivity of $B_{T_{1}}$. $T_{1}$ is related to external
dissipation. The larger $T_{1}$ is, the smaller external photon losses is.

Using Eqs. (\ref{15}) and (\ref{27}), and the Weyl ordering invariance under
similarity transformations, the equivalent parity operator including photon
losses $\Pi _{b}^{loss}$ can be calculated as

\begin{eqnarray}
\Pi _{b}^{loss} &=&\frac{\pi }{2}\left. _{f_{1}}\left\langle 0\right\vert
\begin{array}{c}
\colon \\
\colon%
\end{array}%
B_{T_{1}}^{\dagger }\delta \left( b\right) \delta \left( b^{\dag }\right)
B_{T_{1}}%
\begin{array}{c}
\colon \\
\colon%
\end{array}%
\left\vert 0\right\rangle _{f_{1}}\right.  \notag \\
&=&\frac{\pi }{2}\left. _{f_{1}}\left\langle 0\right\vert
\begin{array}{c}
\colon \\
\colon%
\end{array}%
\delta (\sqrt{T_{1}}b+\sqrt{1-T_{1}}f_{1})\right.  \notag \\
&&\times \delta \left( \sqrt{T_{1}}b^{\dag }+\sqrt{1-T_{1}}f_{1}^{\dag
}\right)
\begin{array}{c}
\colon \\
\colon%
\end{array}%
\left\vert 0\right\rangle _{f_{1}},  \label{28}
\end{eqnarray}%
where $\left\vert 0\right\rangle _{f_{1}}$ is vacuum noise inputting BS $%
B_{T_{1}}$. In a similar way to deriving Eq. (\ref{22}), using Eqs. (\ref{19}%
)-(\ref{21}), the normal ordering form of $\Pi _{b}^{loss}$ is derived as%
\begin{equation}
\Pi _{b}^{loss}=\colon \exp \left( -2T_{1}b^{\dagger }b\right) \colon .
\label{29}
\end{equation}%
It is clear that for the case of $T_{1}=1$ corresponding to the photon
lossless, $\Pi _{b}^{loss}$ just reduces to $\Pi _{b}=\colon \exp \left(
-2b^{\dagger }b\right) \colon ,$ as expected. Thus, the expectation value of
parity detection in the case of photon loss can be transformed to be $%
\left\langle \Pi _{b}^{loss}\right\rangle =\left\langle \text{out}%
\right\vert \Pi _{b}^{loss}\left\vert \text{out}\right\rangle $, where $%
\left\vert \text{out}\right\rangle $ is the output state after the second BS
of MZI and before the photon loss.

In our scheme, using the completeness relation of coherent states, and
combining the unitary transformations $U_{MZI}a^{\dagger }U_{MZI}^{\dag
}=a^{\dagger }\cos \frac{\varphi }{2}+b^{\dagger }\sin \frac{\varphi }{2}$, $%
U_{MZI}b^{\dagger }U_{MZI}^{\dag }=b^{\dagger }\cos \frac{\varphi }{2}%
-a^{\dagger }\sin \frac{\varphi }{2}$ as well as the relation $%
U_{MZI}\left
\vert 0,0\right \rangle =\left \vert 0,0\right \rangle $, the
output state can be shown as%
\begin{eqnarray}
\left \vert \text{out}\right \rangle &=&\frac{\text{sech}r}{n!}\frac{%
\partial ^{2n}}{\partial x^{n}\partial y^{n}}\int \frac{d^{2}\alpha
d^{2}\beta }{\pi ^{2}}  \notag \\
&&\times \exp [-\left \vert \alpha \right \vert ^{2}-\left \vert \beta
\right \vert ^{2}+\alpha ^{\ast }\beta ^{\ast }\tanh r  \notag \\
&&-xy\tanh r+\alpha ^{\ast }x\text{sech}r+\beta ^{\ast }y\text{sech}r]
\notag \\
&&\times \exp [(\alpha \cos \frac{\varphi }{2}-\beta \sin \frac{\varphi }{2}%
)a^{\dagger }  \notag \\
&&+(\alpha \sin \frac{\varphi }{2}+\beta \cos \frac{\varphi }{2})b^{\dagger
}]\left \vert 00\right \rangle |_{x=y=0}.  \label{30}
\end{eqnarray}%
Thus, by further inserting the completeness relation of coherent states and
using Eq. (\ref{21}),\ the parity measurement in realistic case is
calculated as

\begin{equation}
\left \langle \Pi _{b}^{loss}\right \rangle =C_{1}\hat{D}_{n}\exp \left[
C_{2}+C_{3}+C_{4}+C_{5}\right] ,  \label{31}
\end{equation}%
where

\begin{eqnarray}
C_{1} &=&\frac{\text{sech}^{2}r}{\sqrt{\omega _{1}}},  \notag \\
C_{2} &=&\frac{\mu _{1}\varkappa _{1}}{\omega _{1}}\left[ 1-\left( \epsilon
_{1}^{2}+\epsilon _{2}\epsilon _{3}\right) \tanh ^{2}r\right] ,  \notag \\
C_{3} &=&\frac{\mu _{1}^{2}\epsilon _{1}\epsilon _{2}}{\omega _{1}}\tanh r,
\notag \\
C_{4} &=&\frac{\varkappa _{1}^{2}\epsilon _{1}\epsilon _{3}}{\omega _{1}}%
\tanh ^{3}r,  \notag \\
C_{5} &=&\left( \epsilon _{1}x+\epsilon _{3}y\right) \tau \text{sech}^{2}r
\notag \\
&&+\epsilon _{1}\epsilon _{3}\left( \tau \text{sech}r\right) ^{2}\tanh r
\notag \\
&&-\left( t\tau +xy\right) \tanh r,  \label{32}
\end{eqnarray}%
and
\begin{eqnarray}
\omega _{1} &=&\left( \left( \epsilon _{1}^{2}+\epsilon _{2}\epsilon
_{3}\right) \tanh ^{2}r-1\right) ^{2}  \notag \\
&&-4\epsilon _{1}^{2}\epsilon _{2}\epsilon _{3}\tanh ^{4}r,  \notag \\
\mu _{1} &=&\left( \epsilon _{1}x+\epsilon _{3}y\right) \text{sech}r\tanh r
\notag \\
&&+\left( 2\epsilon _{1}\epsilon _{3}\tau \tanh ^{2}r+t\right) \text{sech}r,
\notag \\
\varkappa _{1} &=&\left( \epsilon _{1}^{2}+\epsilon _{2}\epsilon _{3}\right)
\tau \text{sech}r\tanh r  \notag \\
&&+\left( \epsilon _{2}x+\epsilon _{1}y\right) \text{sech}r,  \notag \\
\epsilon _{1} &=&-T_{1}\cos \varphi ,  \notag \\
\epsilon _{2} &=&1-T_{1}\left( 1+\sin \varphi \right) ,  \notag \\
\epsilon _{3} &=&1-T_{1}\left( 1-\sin \varphi \right) ,  \label{33}
\end{eqnarray}%
where $\varphi \longrightarrow \varphi +\pi /2$ is used again. In
particular, when $T_{1}=1$, i.e., the ideal case, Eq. (\ref{31}) can be
simplified to be Eq. (\ref{23}). Furthermore, when $n=0$, Eq. (\ref{31})
becomes $\left \langle \Pi _{b}^{loss}\right \rangle =\frac{\text{sech}^{2}r%
}{\sqrt{\omega _{0}}},$ as expected. Using the expectation value $%
\left
\langle \Pi _{b}^{loss}\right \rangle $ of parity operator under
external photon losses in Eq. (\ref{31}), we can further obtain the phase
sensitivity $\Delta \varphi $ by using
\begin{equation}
\triangle \varphi =\frac{\triangle \Pi _{b}^{loss}}{\left \vert \partial
\left \langle \Pi _{b}^{loss}\right \rangle /\partial \varphi \right \vert },
\label{34}
\end{equation}%
which is similar to deriving Eq. (\ref{23}).

According to Eq. (\ref{34}), we can further investigate the phase
sensitivity when the Laguerre polynomial excitation squeezed state as input
of MZI. As shown in Fig. 3, for both ideal and realistic cases, the phase
sensitivity $\Delta \varphi $ is plotted as the function of the squeezing
parameter $r$ and the transmissivity $T_{1}$\ of $B_{T_{1}}$ for some given
parameters. From Fig. 3, it is clear that the phase sensitivity $\Delta
\varphi $ in the case of external dissipation is worse than that in the
ideal case ($T_{1}=1$). However, $\Delta \varphi $ can be still improved
with the increase of the excited photon number $n$ for any $r$. In addition,
it is found from Fig. 3(b) that $\Delta \varphi $ can be improved with the
increase of $T_{1}$.

\begin{figure}[tbh]
\label{Fig3} \centering
\subfigure{
\begin{minipage}[b]{0.5\textwidth}
\includegraphics[width=0.83\textwidth]{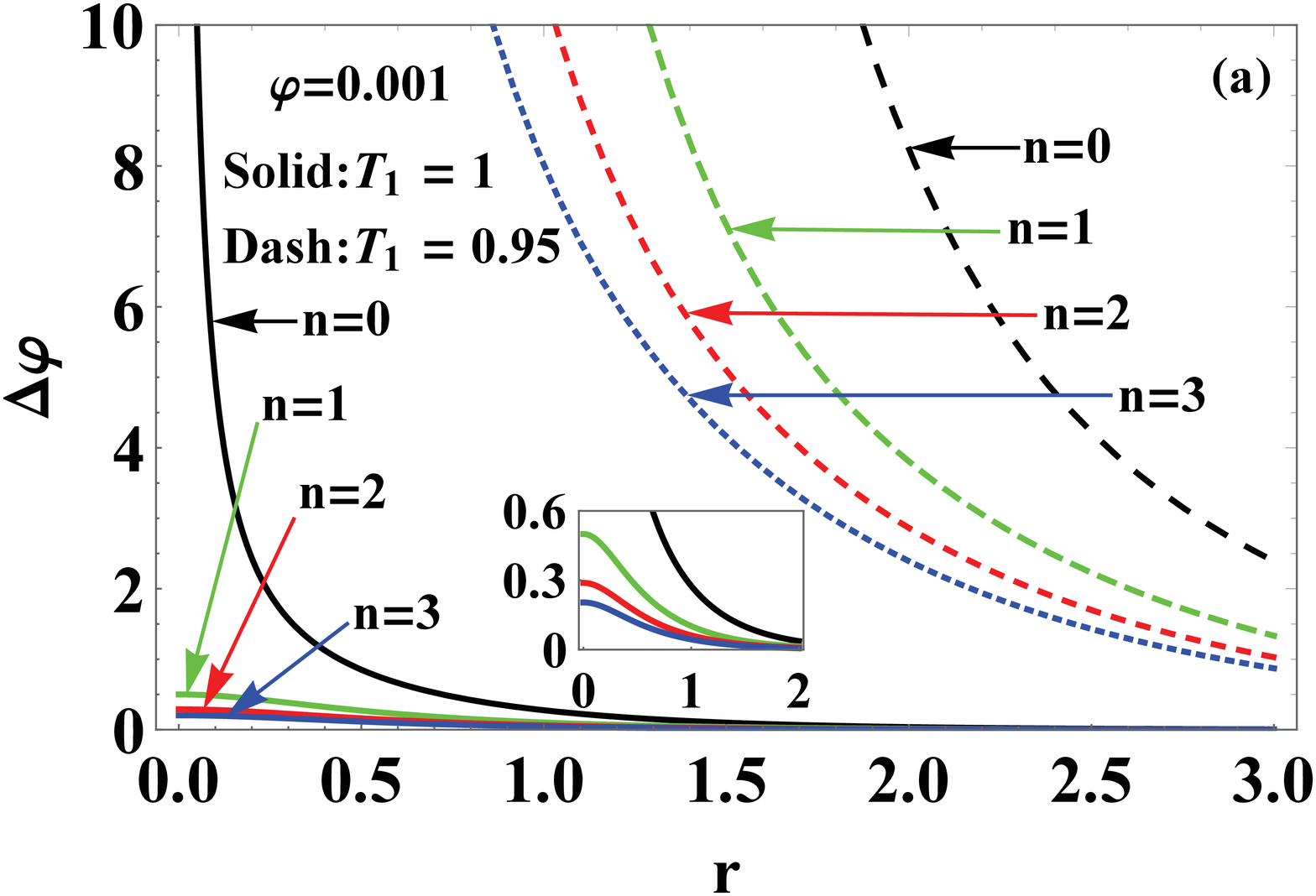}\\
\includegraphics[width=0.83\textwidth]{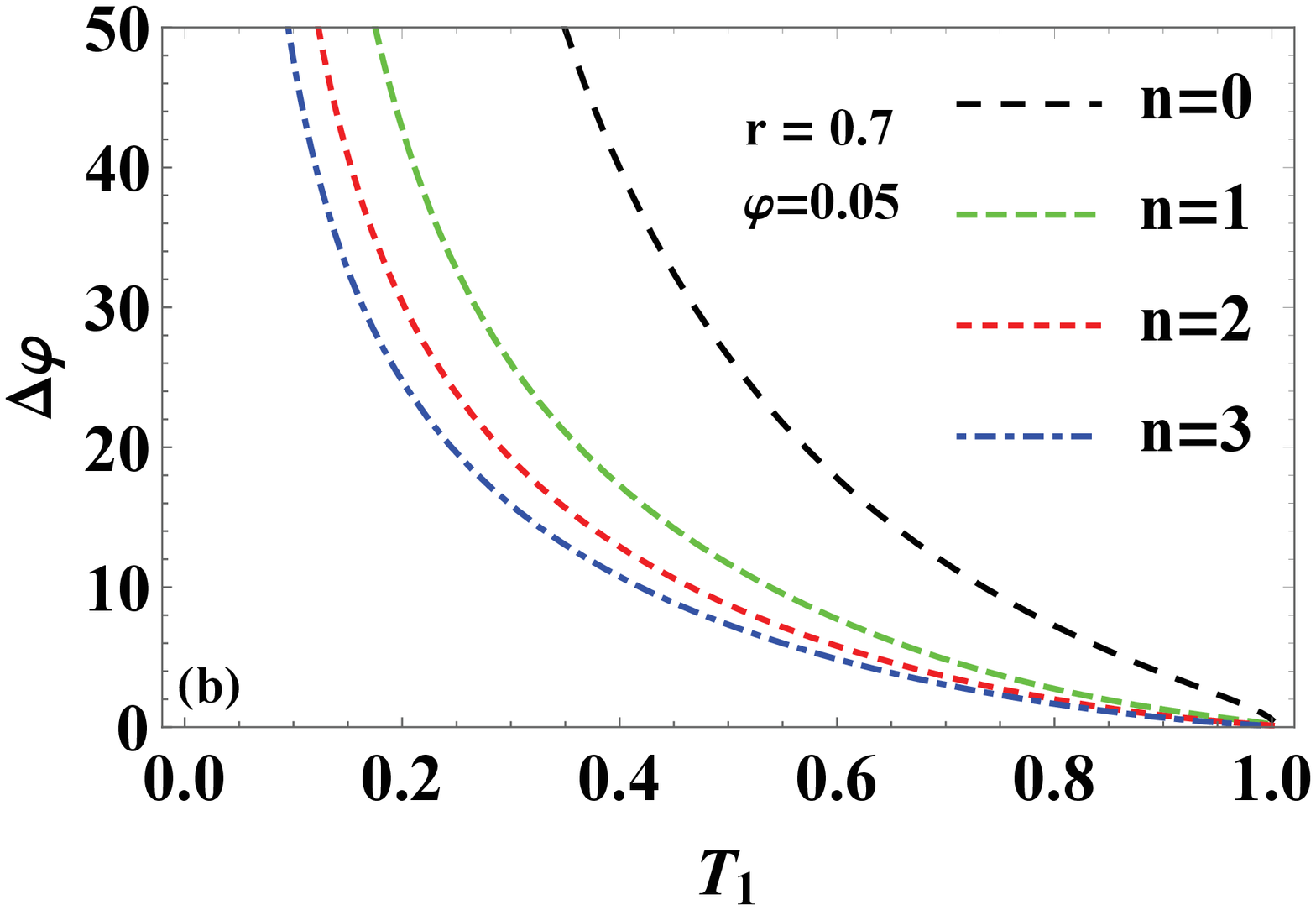}
\end{minipage}}
\caption{For the photon number $n=0,1,2,3$, (a) the phase sensitivity $%
\Delta \protect\varphi $ as a function of the\ squeezing parameters $r$,
for\ the phase shift $\protect\varphi =0.001$, the transmissivity of $%
B_{T_{1}}$ $T_{1}=1$ and $T_{1}=0.95$, (b) for $r=0.7$ and $\protect\varphi %
=0.05$, $\Delta \protect\varphi $ as a function of $T_{1}$.}
\end{figure}

Fig. 4 presents the relation between the phase sensitivity $\Delta \varphi $
and the phase shift $\varphi $ for different excited photon-number $n$ and $%
T_{1}$ as well as given parameter $r=0.7$. From Fig. 4(a), it is shown that
(i) in the ideal case ($T_{1}=1$), the optimal phase sensitivity is at the
point with $\varphi =0$, and it becomes better as $n$ increases. Compared
with the TMSV as inputs, however, the improved region of $\Delta \varphi $
becomes smaller with the increase of $n$. (ii) For the realistic case (say $%
T_{1}=0.95$), the optimal point of the phase sensitivity will deviate from $%
\varphi =0$, and the $\Delta \varphi $ value corresponding to optimal point
decreases with the the increase of $n$. It is interesting to notice that,
even in the realistic case, the phase sensitivity still surpass that by the
TMSV in the ideal case, but the improved region of $\varphi $ becomes
smaller with the increase of $n$ which is similar to the ideal case.

On the other hand, the energy is an important index to measure the phase
sensitivity, here we further consider the phase estimation when fixing the
total initial energy. Fig. 4(b) shows the phase sensitivity $\Delta \varphi $
as the function of $\varphi $ for different $n$ and $T_{1}$ as well as given
total average photon number $\bar{N}=8$. It is found that (i) in the ideal
case ($T_{1}=1$), the optimal phase sensitivity is at the point with $%
\varphi =0$, but it becomes worse as $n$ increases, which is the opposite to
the above situation. However, it is interesting that the improved region of $%
\Delta \varphi $ becomes bigger with the increase of $n$, i.e., the phase
sensitivity is more stable with respect to the phase shift. This implies
that, when fixing the total initial energy, although the optimal phase
sensitivity becomes worse, the improved region will become broader and more
stable. (ii) In the realistic case (say $T_{1}=0.95$), it is clearly seen
that external dissipation causes the optimal phase to deviate from $\varphi
=0$ and the optimal value of $\Delta \varphi $ decreases with the increase
of $n$, i.e., the phase sensitivity becomes higher as $n$ increases which is
similar to the case in Fig. 4(a). In addition, the improved region becomes
bigger as $n$ increases, which is different from the case in Fig. 4(a). In a
word, the phase sensitivity in the realistic case increases with the excited
photon number, whether the initial energy is fixed or not.

\begin{figure}[tbh]
\label{Fig4} \centering
\subfigure{
\begin{minipage}[b]{0.5\textwidth}
\includegraphics[width=0.83\textwidth]{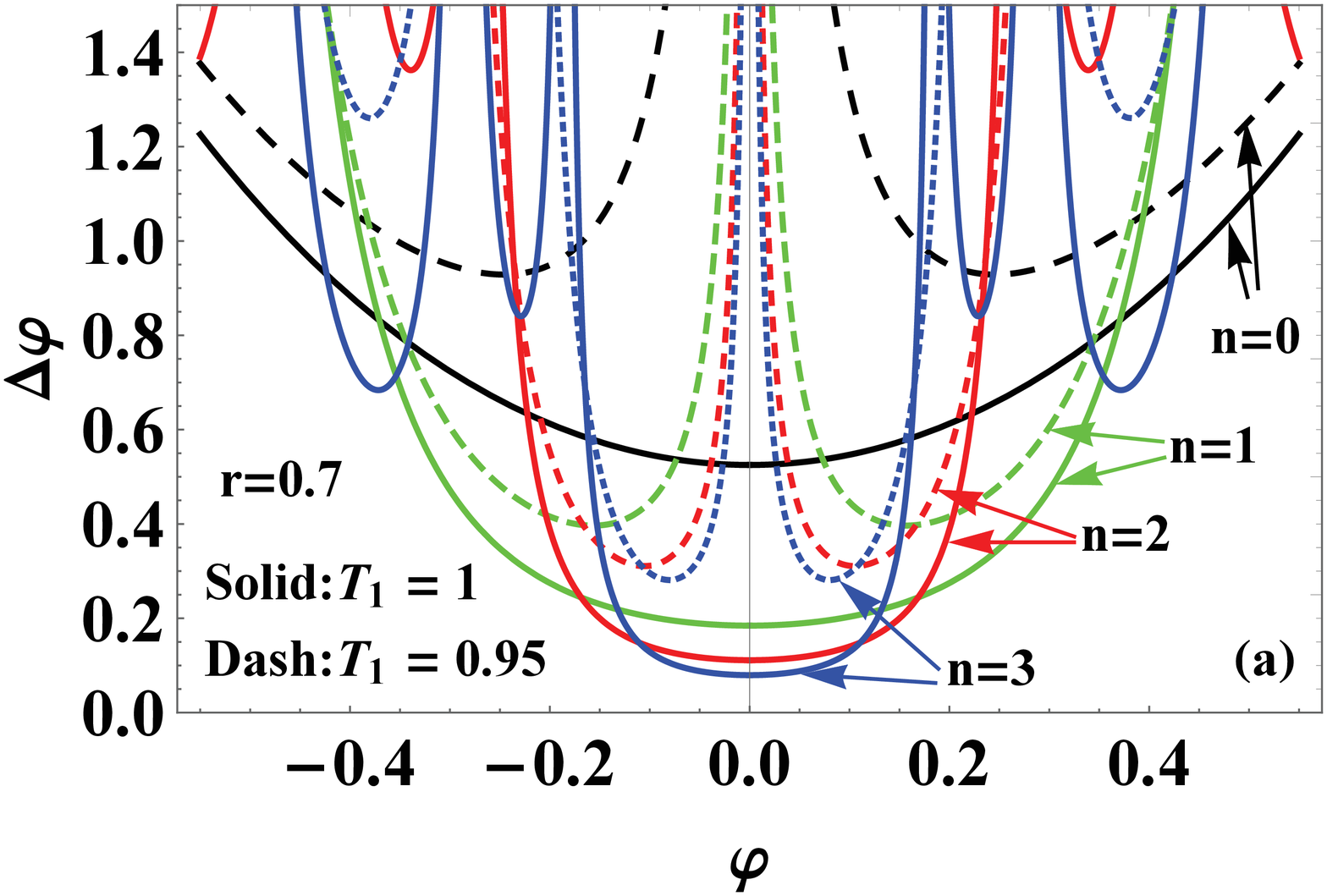}\\
\includegraphics[width=0.83\textwidth]{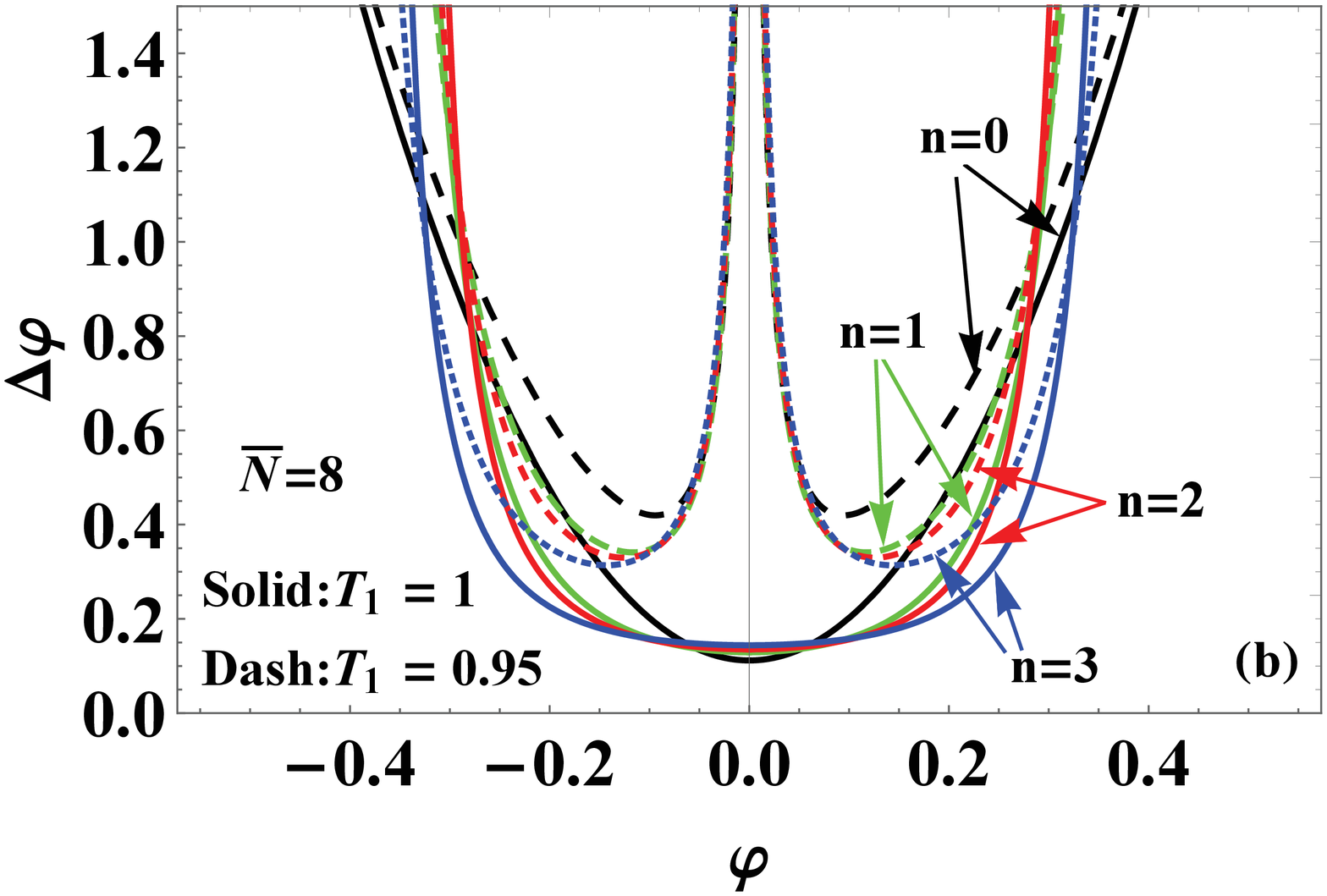}
\end{minipage}}
\caption{The phase sensitivity $\Delta \protect\varphi $ as a function of
the phase shift $\protect\varphi $, for the photon number $n=0,1,2,3$, the
transmissivity of $B_{T_{1}}$ $T_{1}=1$ and $T_{1}=0.95$ (a) Squeezing
parameter $r=0.7$, (b) the total average photon number $\overline{N}=8$.}
\end{figure}

\subsection{Effects of photon losses between phase shifter and BS2 (internal
dissipation)}

In this subsection, we examine the effects of photon losses between phase
shifter and BS2 on the phase sensitivity. We name the photon losses between
them as internal dissipation, as shown in Fig. 2(b). In a similar way, we
adopt an optical BS $B_{T_{2}}$ with a factor $T_{2}$ to simulate the
internal photon-losses process, whose transform relation is
\begin{equation}
B_{T_{2}}^{\dagger }\left(
\begin{array}{c}
b \\
f_{2}%
\end{array}%
\right) B_{T_{2}}=\left(
\begin{array}{cc}
\sqrt{T_{2}} & \sqrt{1-T_{2}} \\
-\sqrt{1-T_{2}} & \sqrt{T_{2}}%
\end{array}%
\right) \left(
\begin{array}{c}
b \\
f_{2}%
\end{array}%
\right) ,  \label{35}
\end{equation}%
where $f_{2}$ ($f_{2}^{\dagger }$) are photon annihilation (creation)
operators corresponding to the dissipative mode $f_{2}$ of $B_{T_{2}}$ and $%
T_{2}$ is the transmissivity of $B_{T_{2}}$. In this case, the average value
of parity detection can be calculated as $\left \langle \widetilde{\Pi }%
_{b}^{loss}\right \rangle =\left \langle \text{in}\right \vert \widetilde{%
\Pi }_{b}^{loss}\left \vert \text{in}\right \rangle $, where $\left \vert
\text{in}\right \rangle $ is the input state of MZI, and $\widetilde{\Pi }%
_{b}^{loss}$ is the equivalent operator of the entire lossy interferometer,
including parity detection, given by
\begin{equation}
\widetilde{\Pi }_{b}^{loss}=\left. _{f_{2}}\left \langle 0\right \vert
B_{1}^{\dagger }U^{\dagger }\left( \varphi \right) B_{T_{2}}^{\dagger
}B_{2}^{\dagger }e^{i\pi b^{\dagger }b}B_{2}B_{T_{2}}U\left( \varphi \right)
B_{1}\left \vert 0\right \rangle _{f_{2}}\right. ,  \label{36}
\end{equation}%
where $B_{1}\left( -\pi /2\right) =e^{-i\frac{\pi }{2}J_{1}}$ and $%
B_{2}\left( \pi /2\right) =e^{i\frac{\pi }{2}J_{1}}$ are BS1 and BS2
operators, respectively, and satisfy the following transform relation:
\begin{eqnarray}
B_{1}^{\dagger }\binom{a}{b}B_{1} &=&\frac{\sqrt{2}}{2}\left(
\begin{array}{cc}
1 & -i \\
-i & 1%
\end{array}%
\right) \binom{a}{b},  \notag \\
B_{2}^{\dagger }\binom{a}{b}B_{2} &=&\frac{\sqrt{2}}{2}\left(
\begin{array}{cc}
1 & i \\
i & 1%
\end{array}%
\right) \binom{a}{b},  \label{37}
\end{eqnarray}%
and $U\left( \varphi \right) =e^{-i\varphi J_{3}}$ is the phase shifter.

In a similar way to deriving Eq. (\ref{28}), by using Eqs. (\ref{35})-(\ref%
{37}) and (\ref{15}), one can obtain the normal ordering of $\widetilde{\Pi }%
_{b}^{loss}$, i.e.,
\begin{equation}
\widetilde{\Pi }_{b}^{loss}=\colon e^{X_{1}a^{\dagger }a-X_{2}b^{\dagger
}a-X_{2}^{\ast }a^{\dagger }b+X_{3}b^{\dagger }b}\colon ,  \label{38}
\end{equation}%
where
\begin{eqnarray}
X_{1} &=&-\frac{2\sqrt{T_{2}}\sin \varphi +1+T_{2}}{2},  \notag \\
X_{2} &=&\frac{\left( T_{2}+1\right) ^{2}-4T_{2}\sin ^{2}\varphi }{2\left(
iT_{2}-i+2\sqrt{T_{2}}\cos \varphi \right) },  \notag \\
X_{3} &=&\frac{2\sqrt{T_{2}}\sin \varphi -1-T_{2}}{2},  \label{39}
\end{eqnarray}%
where $\varphi \longrightarrow \varphi +\pi /2$. In particular, when $%
T_{2}=1 $ corresponding to the ideal case, we have $X_{1}\rightarrow -\sin
\varphi -1,X_{2}=\cos \varphi ,X_{3}=\sin \varphi -1$. Then $\widetilde{\Pi }%
_{b}^{loss}\rightarrow \colon e^{\left( -\sin \varphi -1\right) a^{\dagger
}a-\cos \varphi \left( b^{\dagger }a+a^{\dagger }b\right) +\left( \sin
\varphi -1\right) b^{\dagger }b}\colon $, as expected (reduces to Eq. (\ref%
{22})).

In our scheme, the input state is given by Eq. (\ref{4}). Thus, by using Eqs.
(\ref{4}), (\ref{21}) and (\ref{38}), and inserting completeness relation of
coherent states, we can get the expectation value of $\widetilde{\Pi }%
_{b}^{loss}$ under the input state, which is given by
\begin{equation}
\left \langle \widetilde{\Pi }_{b}^{loss}\right \rangle =D_{1}\hat{D}%
_{n}\left \{ \exp \left[ D_{2}+D_{3}+D_{4}+D_{5}\right] \right \} ,
\label{40}
\end{equation}%
where%
\begin{eqnarray}
D_{1} &=&\frac{\text{sech}^{2}r}{\sqrt{\omega _{2}}},  \notag \\
D_{2} &=&\frac{\mu _{2}\varkappa _{2}\left( 1-E\tanh ^{2}r\right) }{\omega
_{2}},  \notag \\
D_{3} &=&\frac{-\mu _{2}^{2}X_{2}\left( X_{3}+1\right) }{\omega _{2}}\tanh r,
\notag \\
D_{4} &=&\frac{\varkappa _{2}^{2}\left( -X_{1}X_{2}^{\ast }-X_{2}^{\ast
}\right) }{\omega _{2}}\tanh ^{3}r,  \notag \\
D_{5} &=&\left( -X_{2}^{\ast }y+X_{1}x+x\right) t\text{sech}^{2}r  \notag \\
&&+\left( -X_{1}X_{2}^{\ast }-X_{2}^{\ast }\right) t^{2}\text{sech}%
^{2}r\tanh r  \notag \\
&&-xy\tanh r-t\tau \tanh r,  \label{41}
\end{eqnarray}%
and%
\begin{eqnarray}
\omega _{2} &=&\left( 1-E\tanh ^{2}r\right) ^{2}  \notag \\
&&-4\left \vert X_{2}\right \vert ^{2}\left( X_{1}+1\right) \left(
X_{3}+1\right) \tanh ^{4}r,  \notag \\
\mu _{2} &=&(-X_{2}^{\ast }y+X_{1}x+x)\text{sech}r\tanh r  \notag \\
&&-2\left( X_{1}+1\right) X_{2}^{\ast }t\tanh ^{2}r\text{sech}r+\tau \text{%
sech}r,  \notag \\
\varkappa _{2} &=&Et\text{sech}r\tanh r  \notag \\
&&+\left( X_{3}+1\right) y\text{sech}r-X_{2}x\text{sech}r,  \notag \\
E &=&\left \vert X_{2}\right \vert ^{2}+X_{1}X_{3}+X_{3}+X_{1}+1.  \label{42}
\end{eqnarray}%
For the ideal case of $T_{2}=1$, Eq. (\ref{40}) reduces to Eq. (\ref{23}),
as expected. In addition, when $n=0$, Eq. (\ref{40}) becomes $\left \langle
\widetilde{\Pi }_{b}^{loss}\right \rangle =\frac{\text{sech}^{2}r}{\sqrt{%
\omega _{0}}}$. Eq. (\ref{40}) is just the parity signal in the presence of
internal dissipation, and it is ready to obtain the phase sensitivity
combining Eqs. (\ref{34}) and (\ref{40}).

\begin{figure}[tbh]
\label{Fig5} \centering
\subfigure{
\begin{minipage}[b]{0.5\textwidth}
\includegraphics[width=0.83\textwidth]{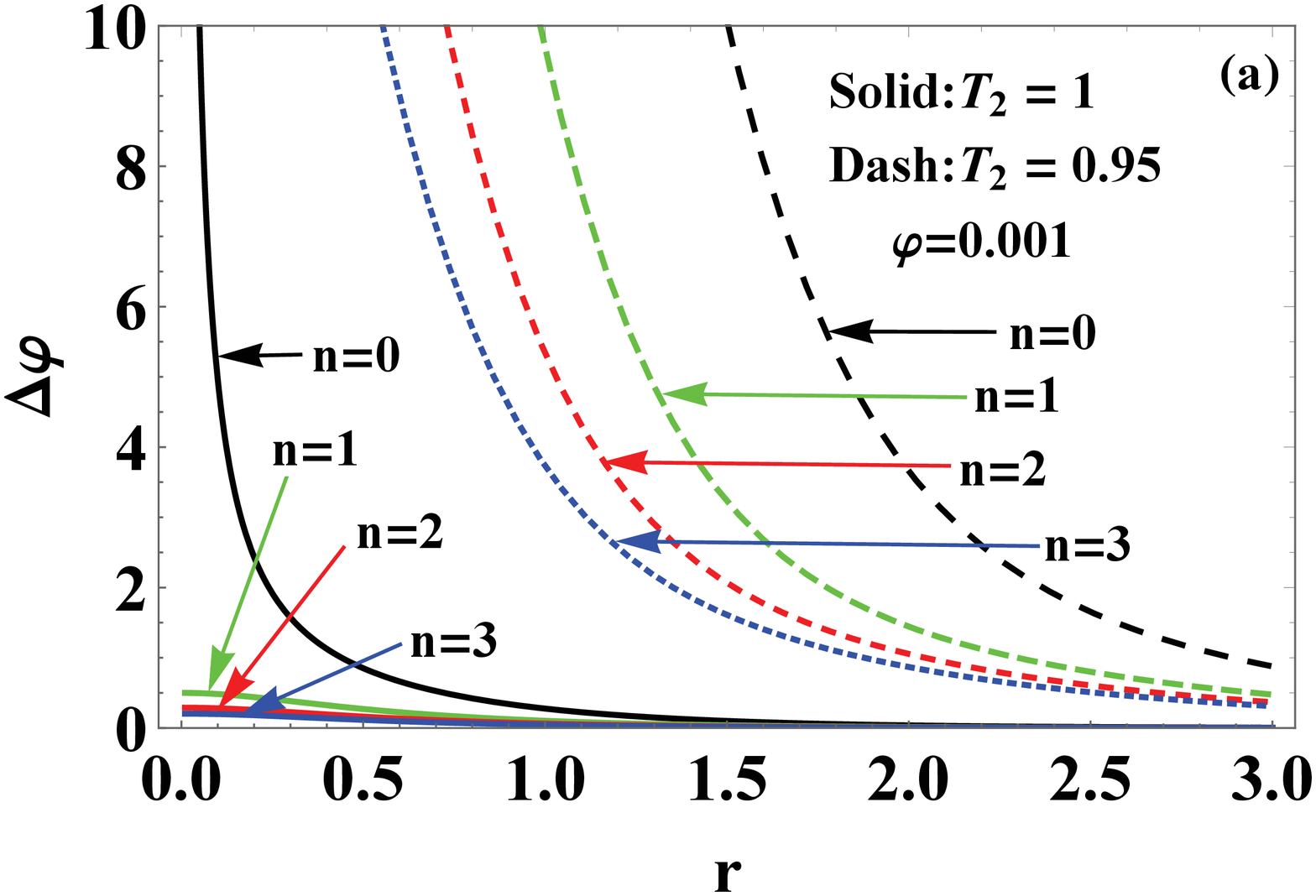}\\
\includegraphics[width=0.83\textwidth]{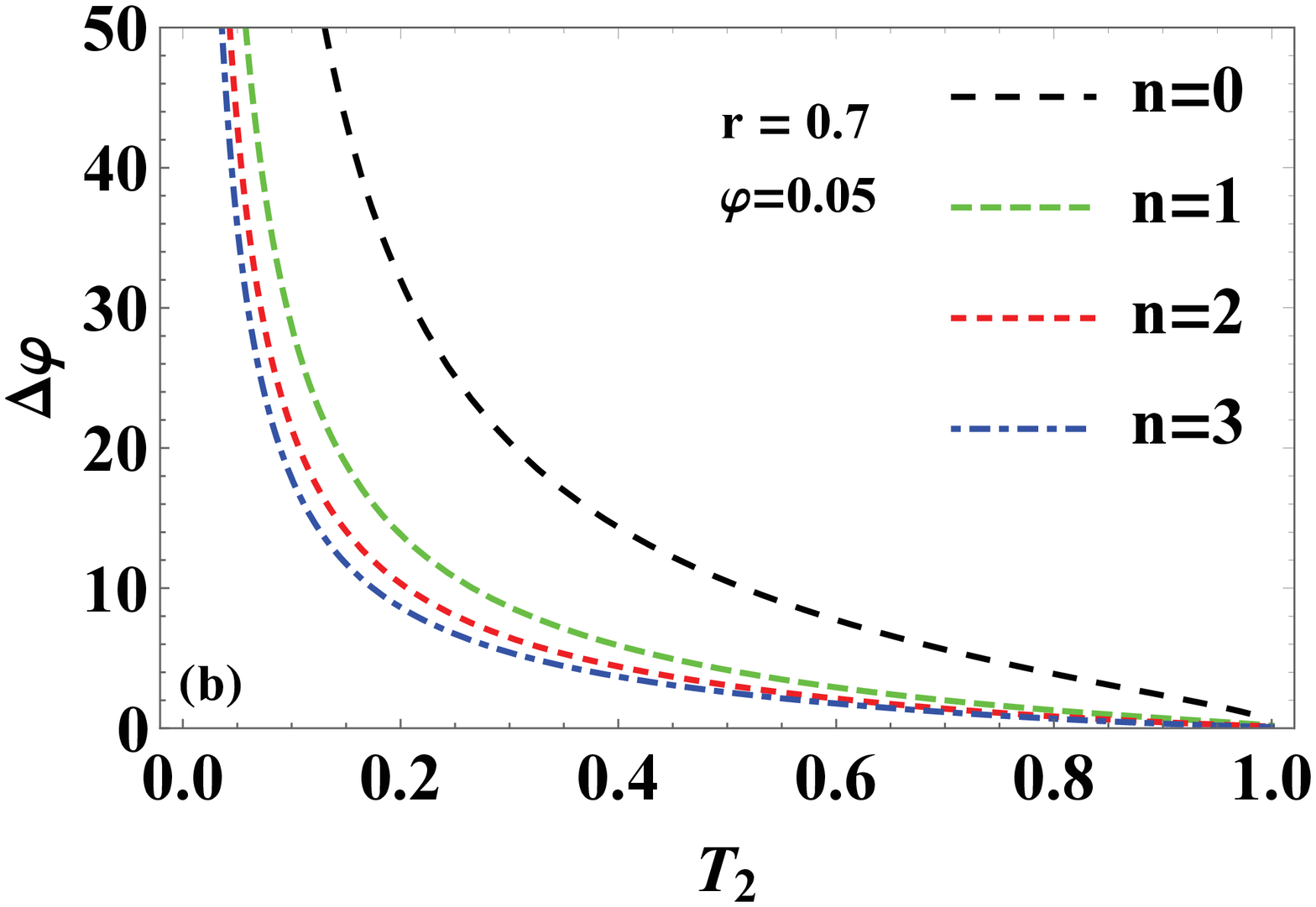}
\end{minipage}}
\caption{For the photon number $n=0,1,2,3$, (a) the phase sensitivity $%
\Delta \protect\varphi $ as a function of the squeezing parameters $r$, for\
the phase shift $\protect\varphi =0.001$, the transmissivity of $B_{T_{2}}$ $%
T_{2}=1$ and $T_{2}=0.95$, (b) for $r=0.7$ and $\protect\varphi =0.05$, $%
\Delta \protect\varphi $ as a function of $T_{2}$.}
\end{figure}

\begin{figure}[tbh]
\label{Fig6} \centering
\subfigure{
\begin{minipage}[b]{0.5\textwidth}
\includegraphics[width=0.83\textwidth]{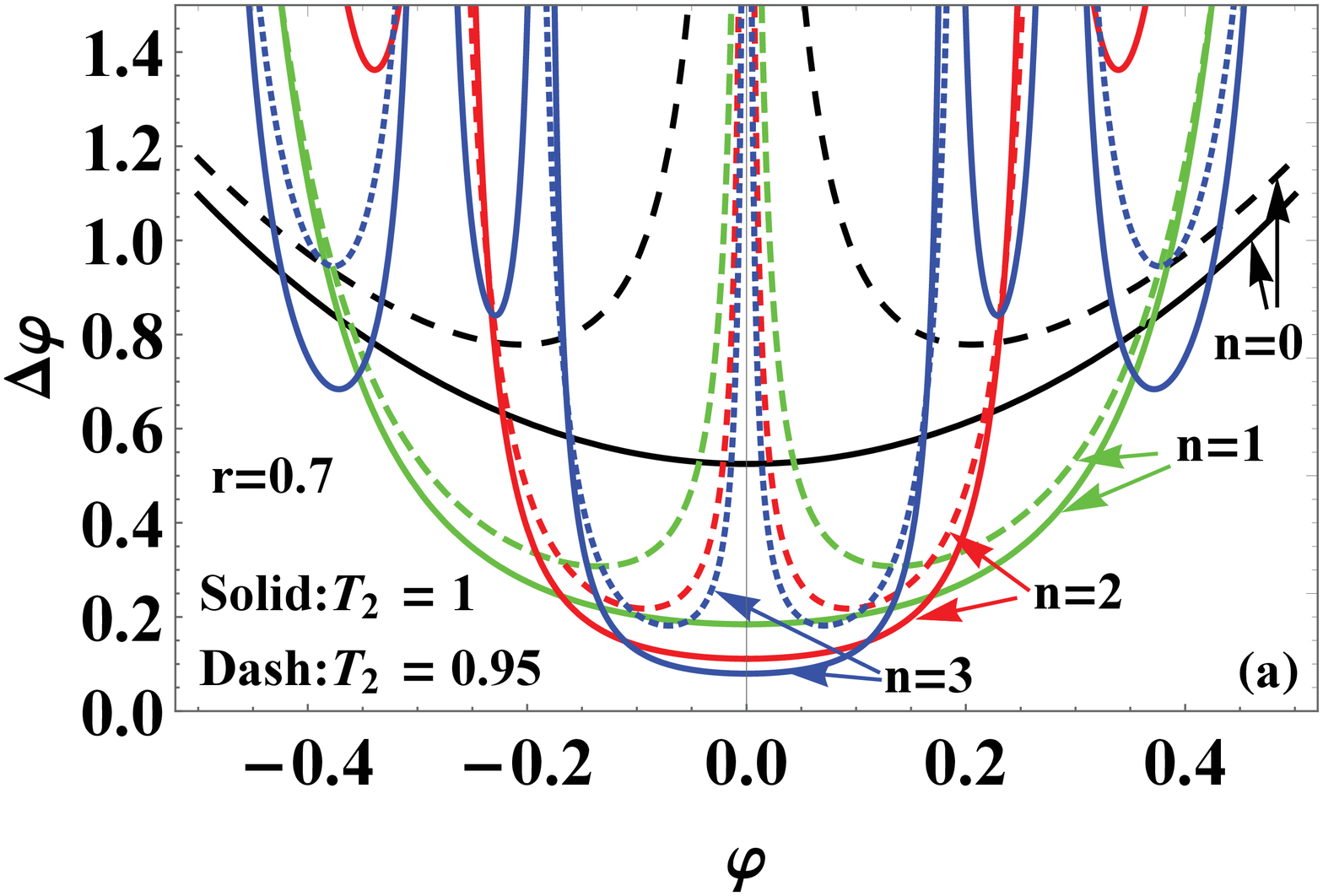}\\
\includegraphics[width=0.83\textwidth]{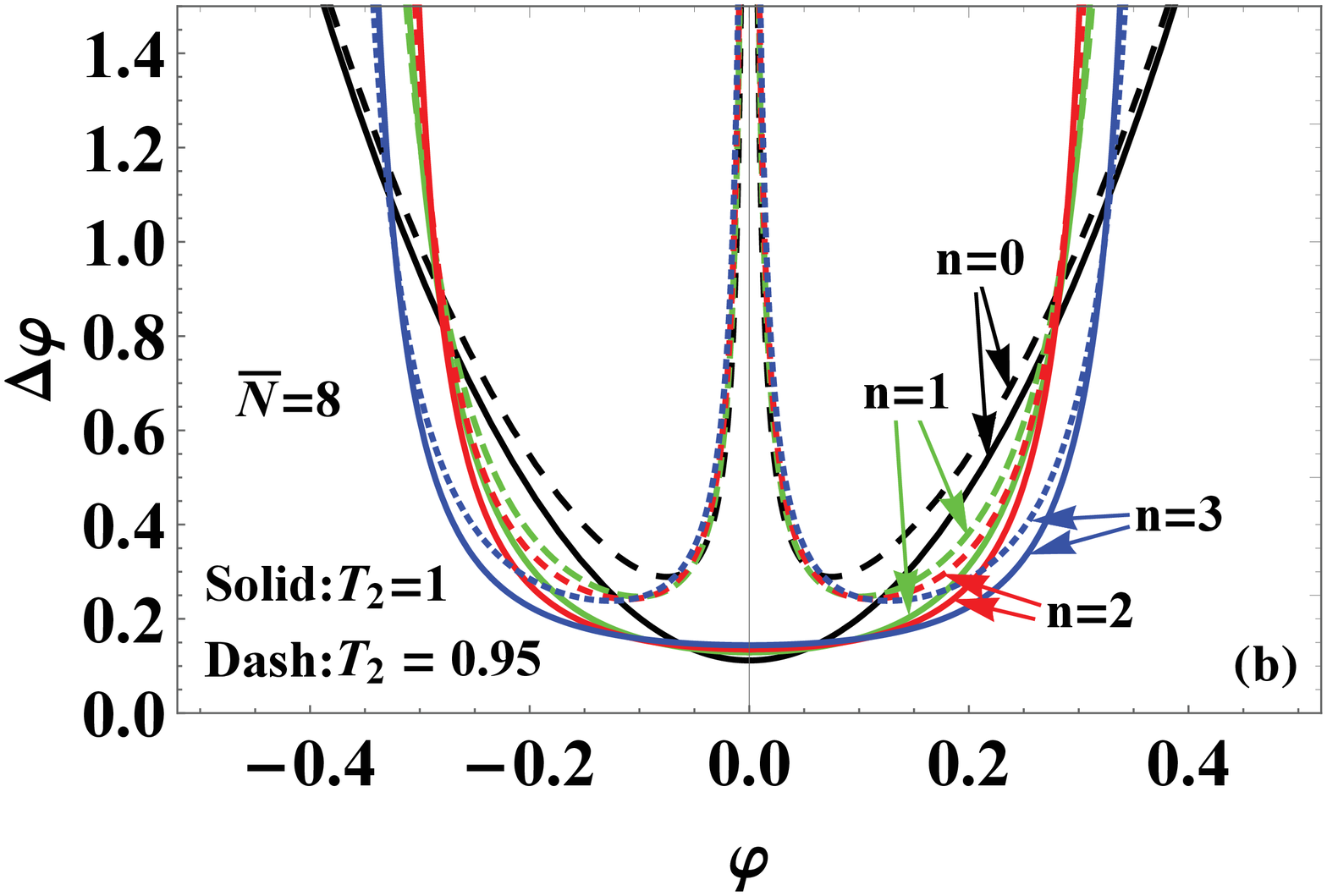}
\end{minipage}}
\caption{The phase sensitivity $\Delta \protect\varphi $ as a function of
the phase shift $\protect\varphi $, for the photon number $n=0,1,2,3$, the
transmissivity of $B_{T_{2}}$ $T_{2}=1$ and $T_{2}=0.95$, (a) for the
squeezing parameter $r=0.7$, (b) for the total average photon number $%
\overline{N}=8$.}
\end{figure}

Similar to Figs. 3 and 4, Figs. 5 and 6 present $\Delta \varphi $ as a
function of squeezing parameter, dissipative factor and phase shift for
other given values. Some similar results can be obtained. Briefly, the phase
sensitivity $\Delta \varphi $ can be enhanced by increasing excited photon
number, and even surpass the ideal phase sensitivity by the TMSV in a
certain region of $\varphi $.

In addition, comparing Figs. 3 and 4 with Figs. 5 and 6, it is ready to see
the difference between internal dissipation and external one on the phase
sensitivity. It is found that the external dissipation has a greater impact
on the phase measurement accuracy than the internal dissipation. To clearly
see this point, at fixed $\varphi =0.05$, $r=0.7$, we give the phase
sensitivity $\Delta \varphi $ as a function of $T_{1}$ ($T_{2}$) for several
different $n=0,1,2,3$ as shown in Fig. 7. This result implies that, to get a
better precision of phase measurement, special attention should be paid to
the control of external photon losses.

\begin{figure}[tph]
\label{Fig7} \centering \includegraphics[width=0.83\columnwidth]{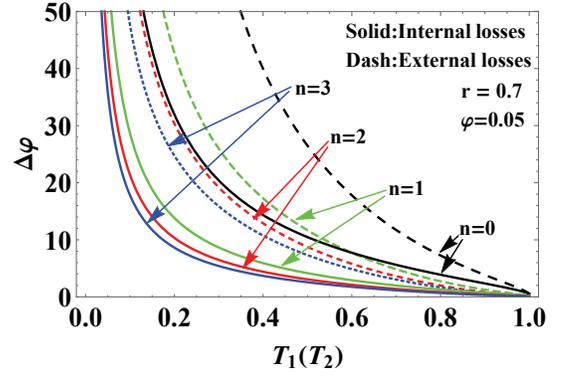}
\caption{{}Comparing the influence of two dissipation ways on the phase
sensitivity $\Delta \protect\varphi $\ for the\ photon number $n=0,1,2,3$.}
\end{figure}

In order to further clearly see the difference between internal and external
dissipations, we plot the phase sensitivity $\Delta \varphi $ as a function
of the squeezing parameter $r$ for different excited photon number $%
n=0,1,2,3 $ (optimized over the parameter $\varphi $) in Fig. 8. Here the
SQL and the HL are also plotted for comparison. From Fig. 8, it is shown
that (i) $\Delta \varphi $ can break the SQL and the HL for $n=0$ [see Fig.
8(a)]. (ii) $\Delta \varphi $ can break through the SQL in a certain range
of $r$. In particular, $\Delta \varphi $ with the internal dissipation can
break through the\ SQL in a larger squeezing region than that with the
external dissipation.

\begin{figure*}[tbh]
\label{Fig8} \centering
\subfigure{
\begin{minipage}[b]{0.83\textwidth}
\includegraphics[width=0.5\textwidth]{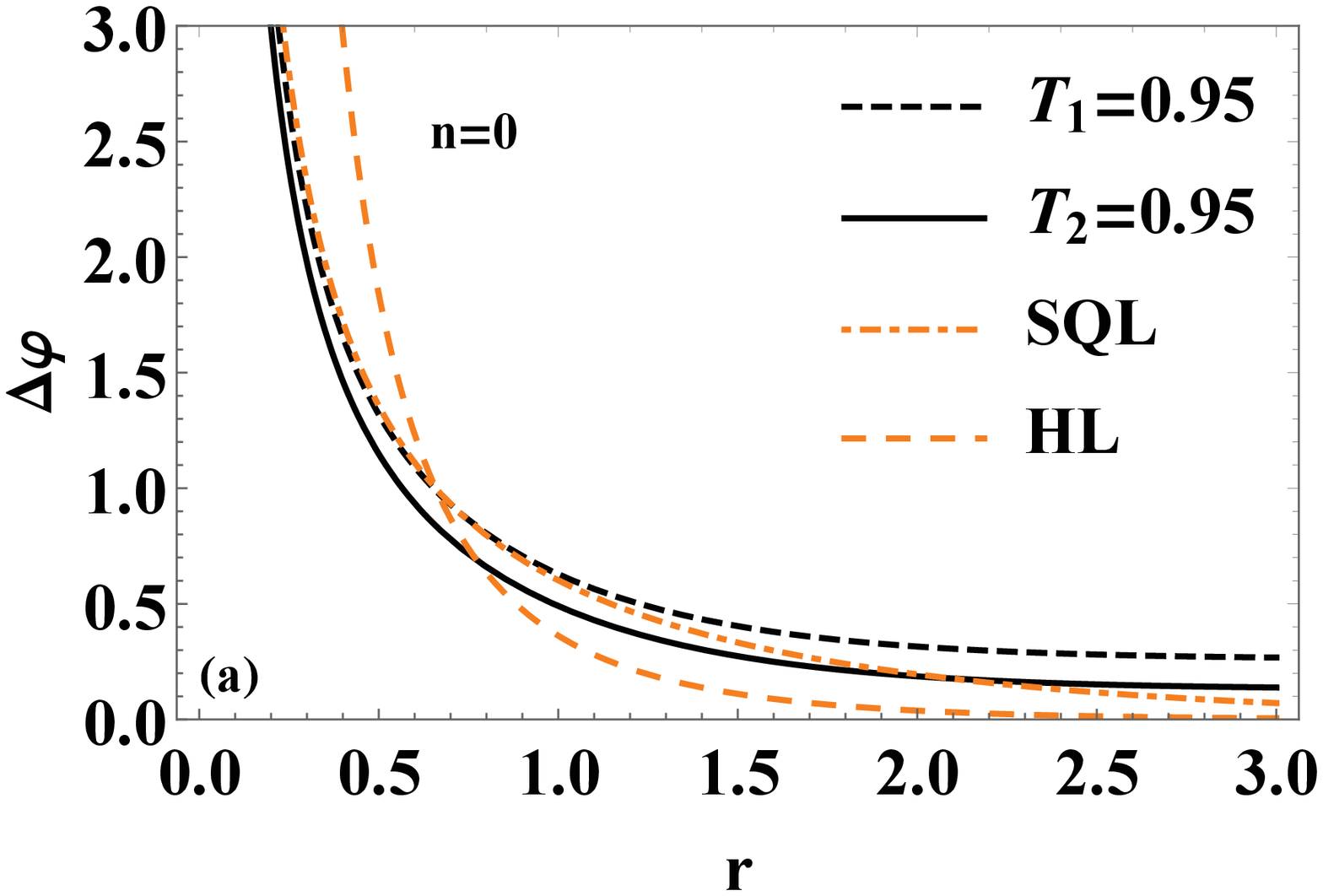}\includegraphics[width=0.5\textwidth]{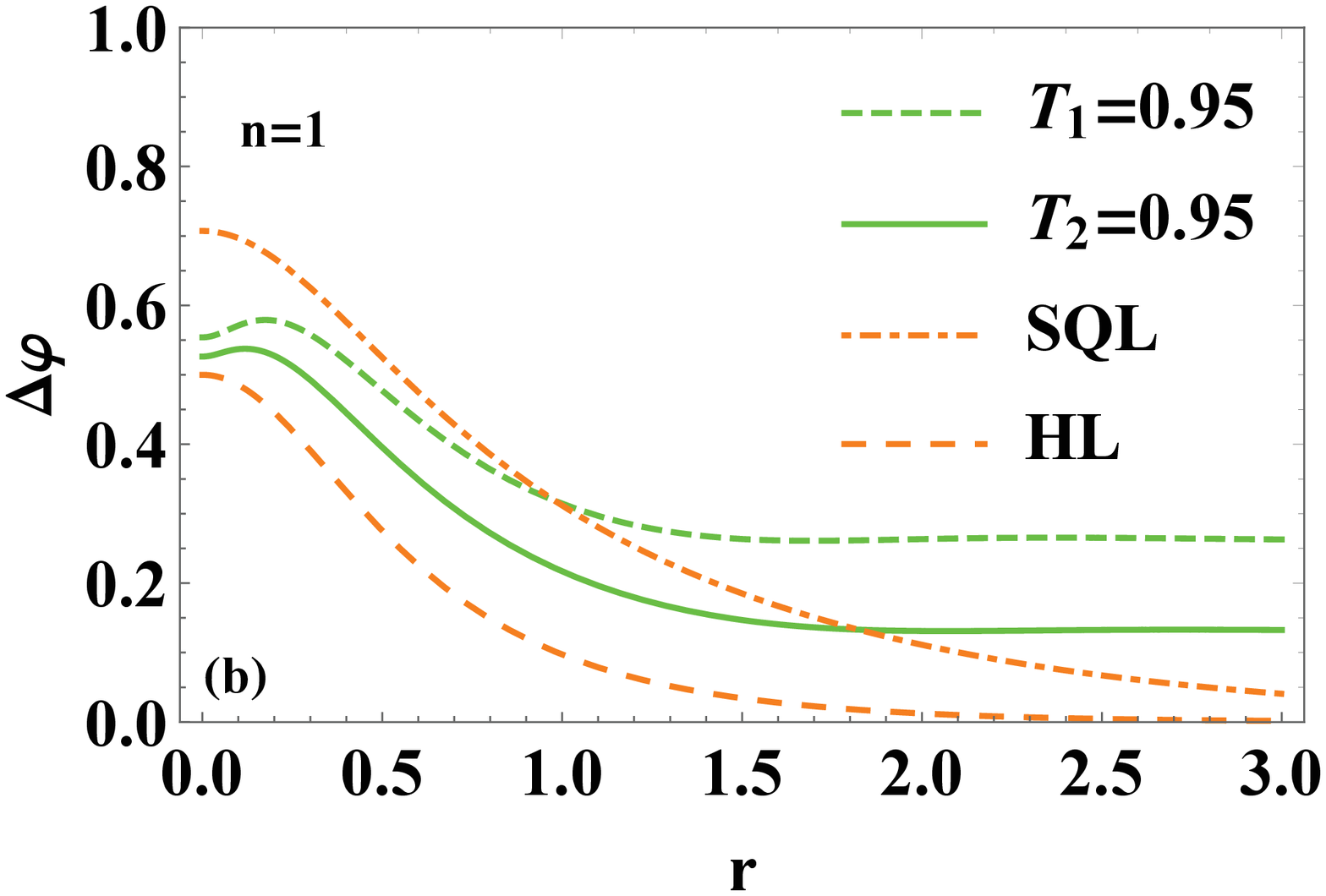}\\
\includegraphics[width=0.5\textwidth]{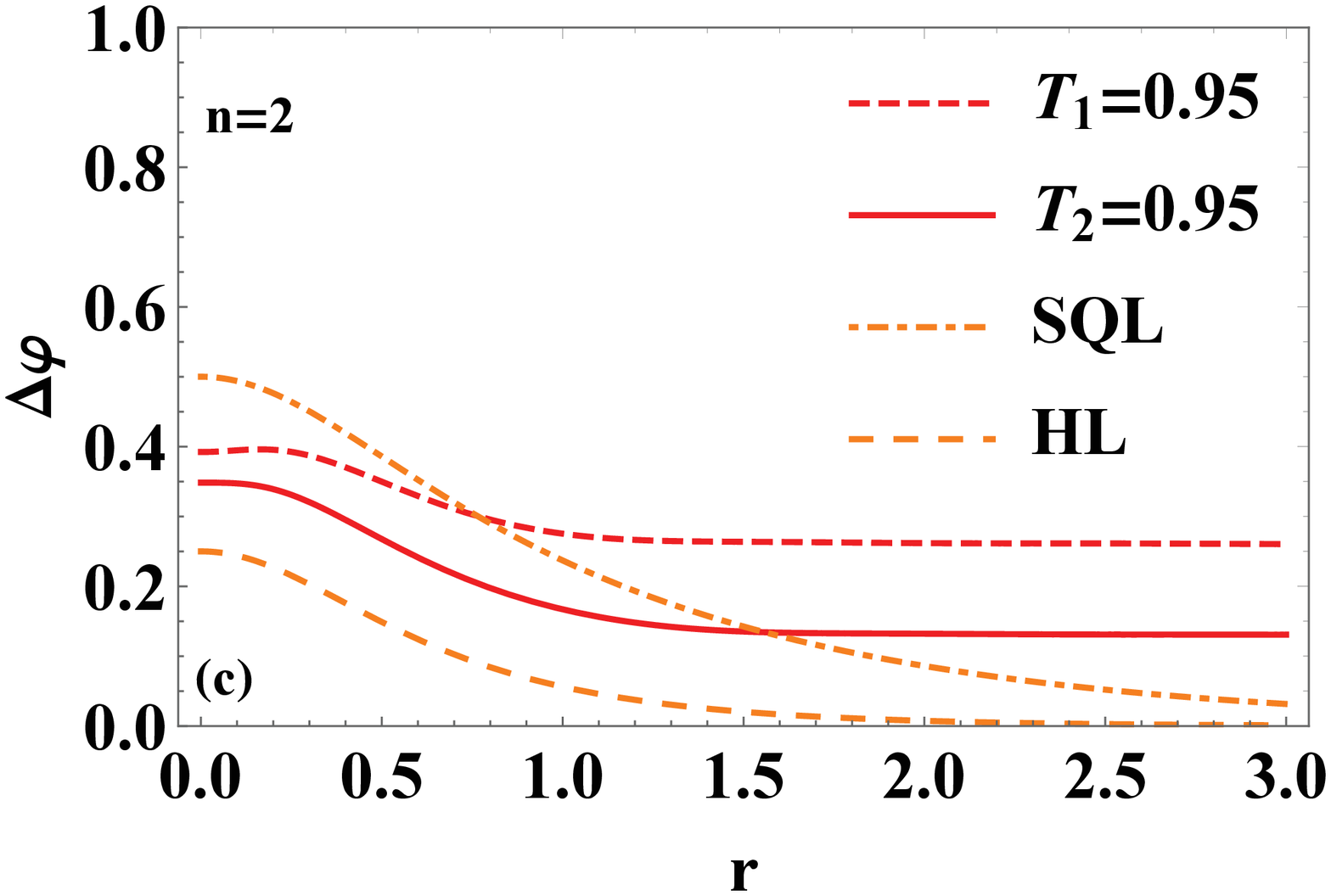}\includegraphics[width=0.5\textwidth]{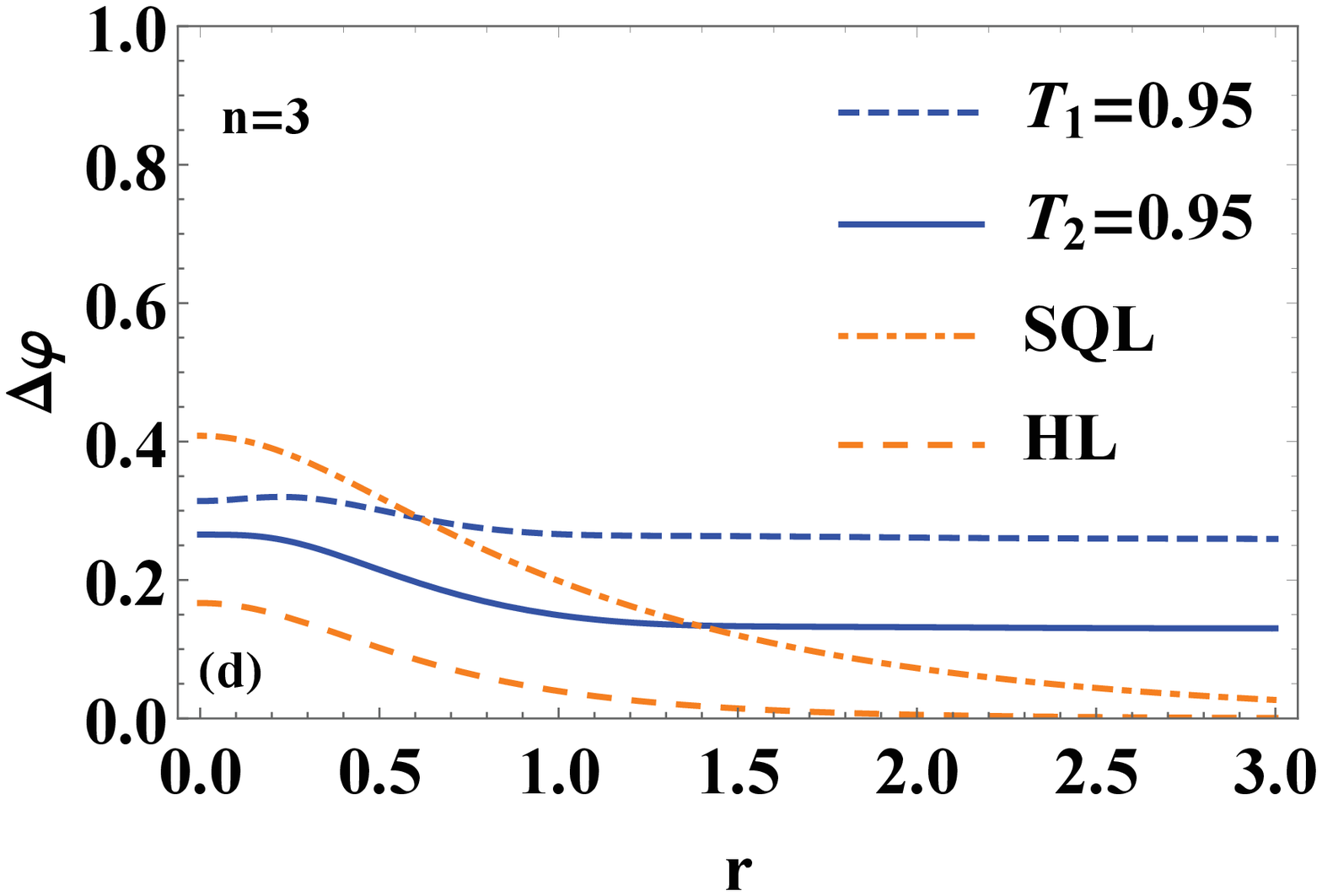}
\end{minipage}}
\caption{The phase sensitivity $\Delta \protect\varphi $ as a function of
the squeezing parameters $r$ in the case of photon losses comparing with the
SQL and the HL (a) for $\protect\varphi =0.2$, $n=0$ and $T_{1}$ or $%
T_{2}=0.96$. (b) for $\protect\varphi =0.15$, $n=1$ and $T_{1}$ or $%
T_{2}=0.95$. (c) for $\protect\varphi =0.12$, $n=2$ and $T_{1}$ or $%
T_{2}=0.95$. (d) for $\protect\varphi =0.1$, $n=3$ and $T_{1}$ or $%
T_{2}=0.95 $.}
\end{figure*}

\section{Effects of photon losses on the QFI}

The QFI theoretically gives the optimal accuracy of phase estimation
independent of special measures, but this optimal accuracy is also affected
by the photon losses in the realistic environment. In this section, we
mainly consider the influences of photon losses along the path of photon
interferometer on the QFI. As shown in Fig. 9, for simplicity, we assume
that photon losses exist in the optical path of mode $b$, and the sources of
photon losses is mainly located before and after the phase shift, which are
respectively simulated by two optical BSs of $B_{\eta _{1}}$ and $B_{\eta
_{2}}$, where $\eta _{1}=\eta _{2}=\eta $ are transmissivities of $B_{\eta
_{1}}$ and $B_{\eta _{2}}$,\ related to the dissipation factor of photon
losses.\

\begin{figure}[tph]
\label{Fig9} \centering \includegraphics[width=0.83\columnwidth]{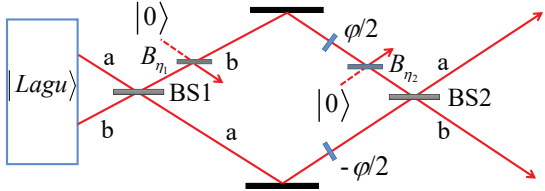}
\caption{{}Schematic diagram of a lossy interferometer. The losses of mode $%
b $ in the interferometer are modeled by adding the fictitious beam
splitters before and after the phase shift.}
\end{figure}

According to the research on the bounds for error estimation in noisy
systems of Escher \emph{et al}. \cite{50}, in this case, the QFI $F_{Q}$\
can be calculated by the following equation:
\begin{equation}
F_{Q}\leq C_{Q}=4\left[ \left \langle \psi \right \vert \hat{H}_{1}\left
\vert \psi \right \rangle -\left \vert \left \langle \psi \right \vert \hat{H%
}_{2}\left \vert \psi \right \rangle \right \vert ^{2}\right] ,  \label{43}
\end{equation}
where the state $\left\vert \psi \right\rangle =e^{-i\frac{\pi }{2}%
J_{1}}\left\vert \text{Lagu}\right\rangle $ is the correlated probe state
after the input state $\left\vert \text{Lagu}\right\rangle $ is injected
into the first optical BS (BS1) of MZI, and Hermitian operators $\hat{H}%
_{1,2}$ are defined by
\begin{eqnarray}
\hat{H}_{1} &=&\sum \limits_{l}\frac{d\hat{\Pi}_{l}^{\dagger }\left( \varphi
\right) }{d\varphi }\frac{d\hat{\Pi}_{l}\left( \varphi \right) }{d\varphi },
\notag \\
\hat{H}_{2} &=&i\sum \limits_{l}\frac{d\hat{\Pi}_{l}^{\dagger }\left(
\varphi \right) }{d\varphi }\hat{\Pi}_{l}\left( \varphi \right) ,  \label{44}
\end{eqnarray}%
where $\hat{\Pi}_{l}\left( \varphi \right) $ are Kraus operators, i.e.,
\begin{equation}
\hat{\Pi}_{l}\left( \varphi \right) =\sqrt{\frac{\left( 1-\eta \right) ^{l}}{%
l!}}e^{-i\varphi \left( \frac{a^{\dagger }a-b^{\dagger }b}{2}+\frac{\gamma l%
}{2}\right) }\eta ^{\frac{b^{\dagger }b}{2}}b^{l}.  \label{45}
\end{equation}%
where $\gamma =0$ and $\gamma =-1$ represent the photon losses before and
after the phase shifter, respectively. $\eta $ is related to the dissipation
factor with $\eta $ $=1$ and $\eta $ $=0$ being the cases of complete
lossless and absorption, respectively.

By combining Eqs. (\ref{43})-(\ref{45}) for further calculation, we can get
\begin{eqnarray}
C_{Q} &=&\left \langle \Delta ^{2}\hat{n}_{a}\right \rangle +\left( \eta
+\gamma \eta -\gamma \right) ^{2}\left \langle \Delta ^{2}\hat{n}_{b}\right
\rangle  \notag \\
&&-2\left( \eta +\gamma \eta -\gamma \right) Cov\left[ \hat{n}_{a},\hat{n}%
_{b}\right]  \notag \\
&&+\left( 1+\gamma \right) ^{2}\eta \left( 1-\eta \right) \left \langle \hat{%
n}_{b}\right \rangle ,  \label{46}
\end{eqnarray}%
where $\hat{n}_{a}=a^{\dagger }a$, $\hat{n}_{b}=b^{\dagger }b$, $%
\left
\langle \Delta ^{2}\hat{n}_{i}\right \rangle =\left \langle \hat{n}%
_{i}^{2}\right \rangle -\left \langle \hat{n}_{i}\right \rangle ^{2}$ ($%
i=a,b $), and $Cov\left[ \hat{n}_{a},\hat{n}_{b}\right] =\left \langle \hat{n%
}_{a}\hat{n}_{b}\right \rangle -\left \langle \hat{n}_{a}\right \rangle
\left
\langle \hat{n}_{b}\right \rangle $ ($\left \langle \cdot
\right
\rangle =\left \langle \psi \right \vert \cdot \left \vert \psi
\right
\rangle $). Minimizing over the parameter $\gamma $ in Eq. (\ref%
{46}) will lead to the minimum value of $C_{Q}$ in the presence of photon
losses, where the optimal value of $\gamma $ can be obtained
\begin{equation}
\gamma _{opt}=\frac{\eta \left \langle \Delta ^{2}\hat{n}_{b}\right \rangle
-Cov\left[ \hat{n}_{a},\hat{n}_{b}\right] -\eta \left \langle \hat{n}%
_{b}\right \rangle }{\left( 1-\eta \right) \left \langle \Delta ^{2}\hat{n}%
_{b}\right \rangle +\eta \left \langle \hat{n}_{b}\right \rangle }.
\label{47}
\end{equation}%
Thus substituting Eq. (\ref{47}) into Eq. (\ref{46}), the minimum value of $%
C_{Q}$ can be ready to obtian.

Using Eqs. (\ref{1}) and (\ref{3}) and the transform relations

\begin{eqnarray}
e^{i\frac{\pi }{2}J_{1}}ae^{-i\frac{\pi }{2}J_{1}} &=&\frac{\sqrt{2}}{2}%
\left( a-ib\right) ,  \notag \\
e^{i\frac{\pi }{2}J_{1}}be^{-i\frac{\pi }{2}J_{1}} &=&\frac{\sqrt{2}}{2}%
\left( b-ia\right) ,  \label{48}
\end{eqnarray}%
one can obtain

\begin{eqnarray}
\left\langle \hat{n}_{a}\right\rangle &=&\left\langle \hat{n}%
_{b}\right\rangle =n\cosh ^{2}r+(n+1)\sinh ^{2}r,  \notag \\
\left\langle \hat{n}_{a}^{2}\right\rangle &=&\left\langle \hat{n}%
_{b}^{2}\right\rangle =\left( 3n^{2}+n\right) \frac{\cosh ^{4}r}{2}  \notag
\\
&&+\left( 3n^{2}+5n+2\right) \frac{\sinh ^{4}r}{2}  \notag \\
&&+\left( 3n^{2}+3n+1\right) \frac{\sinh ^{2}\left( 2r\right) }{2},
\label{49}
\end{eqnarray}%
and%
\begin{eqnarray}
\left\langle \hat{n}_{a}\hat{n}_{b}\right\rangle &=&\left( n^{2}-n\right)
\frac{\cosh ^{4}r}{2}  \notag \\
&&+\left( n^{2}+3n+2\right) \frac{\sinh ^{4}r}{2}  \notag \\
&&+\left( n^{2}+n\right) \frac{\sinh ^{2}\left( 2r\right) }{2}.  \label{50}
\end{eqnarray}%
By combining Eq. (\ref{46}) with Eq. (\ref{47}) and further using Eq. (\ref%
{49}) and Eq. (\ref{50}), we can get the value of $\gamma _{opt}$ and the
minimum value of $C_{Q}$, i.e. the QFI $F_{Q}$\ for the phase shift in the
presence of photon losses in MZI, not shown here for simplicity. In
particular, for the case of the transmissivity $\eta =1$ corresponding to
the ideal case, the expression for $F_{Q}$ just reduces to Eq. (\ref{11}),
as expected.

\begin{figure}[tph]
\label{Fig10} \centering \includegraphics[width=0.83\columnwidth]{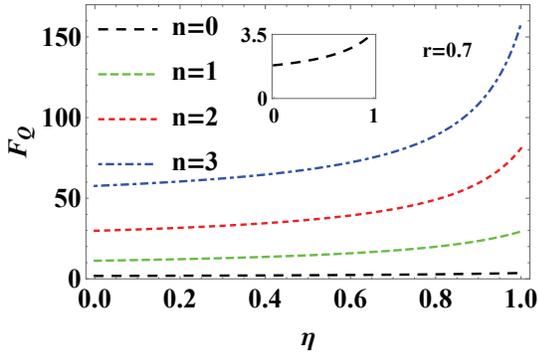}
\caption{{}The quantum Fisher information $F_{Q}$ as a function of the
transmissivity $\protect\eta $, for the photon number $n=0,1,2,3$ and the
squeezing parameter $r=0.7$.}
\end{figure}

\begin{figure}[tbh]
\label{Fig11} \centering
\subfigure{
\begin{minipage}[b]{0.5\textwidth}
\includegraphics[width=0.83\textwidth]{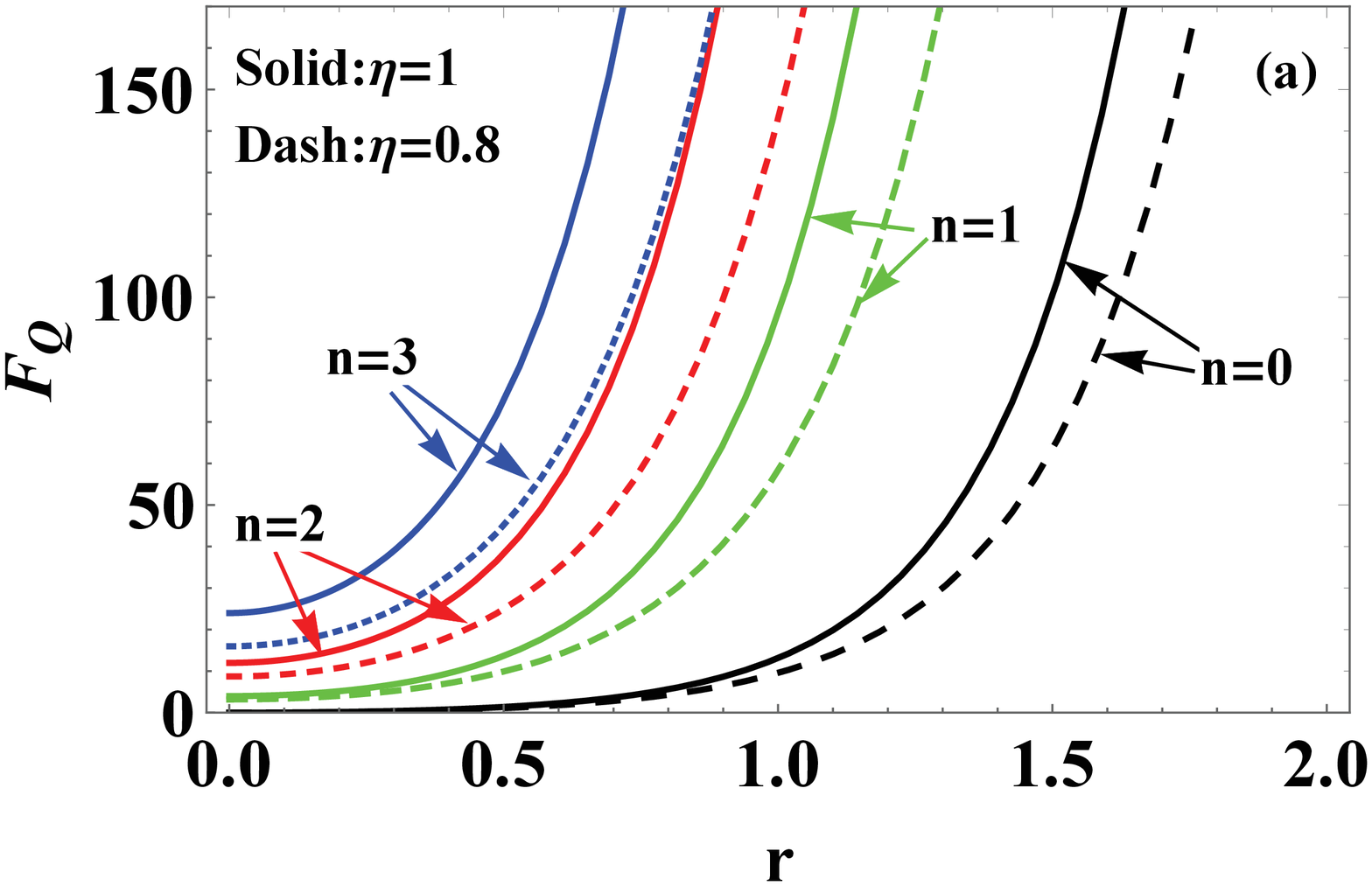}\\
\includegraphics[width=0.83\textwidth]{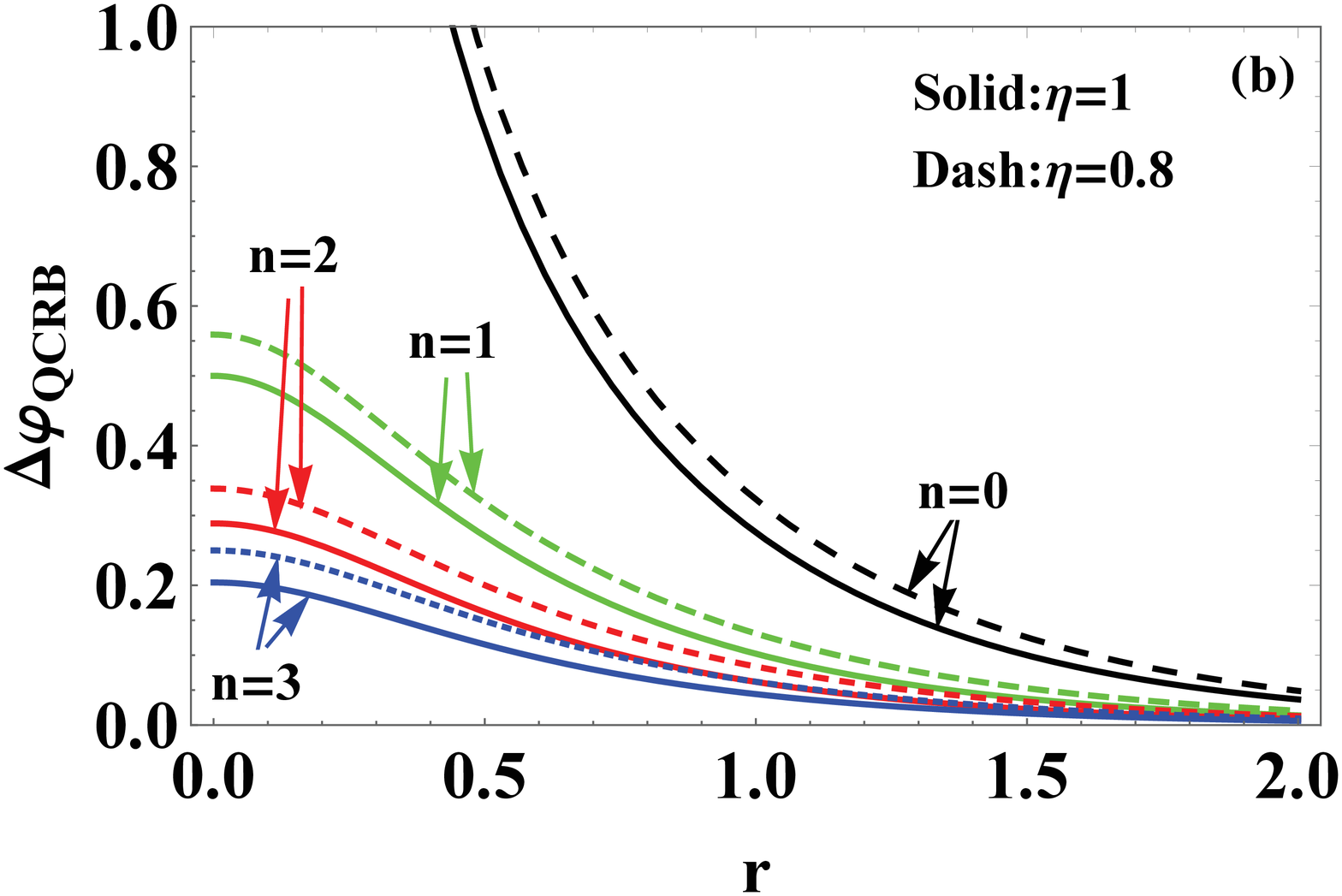}
\end{minipage}}
\caption{For the photon number $n=0,1,2,3$, the transmissivity $\protect\eta %
=1$ and $\protect\eta =0.8$, (a) the quantum Fisher information $F_{Q}$ as a
function of the squeezing parameter $r$, (b) $\Delta \protect\varphi _{QCRB}$
as a function of $r$.}
\end{figure}

\begin{figure*}[tbh]
\label{Fig12} \centering
\subfigure{
\begin{minipage}[b]{0.83\textwidth}
\includegraphics[width=0.5\textwidth]{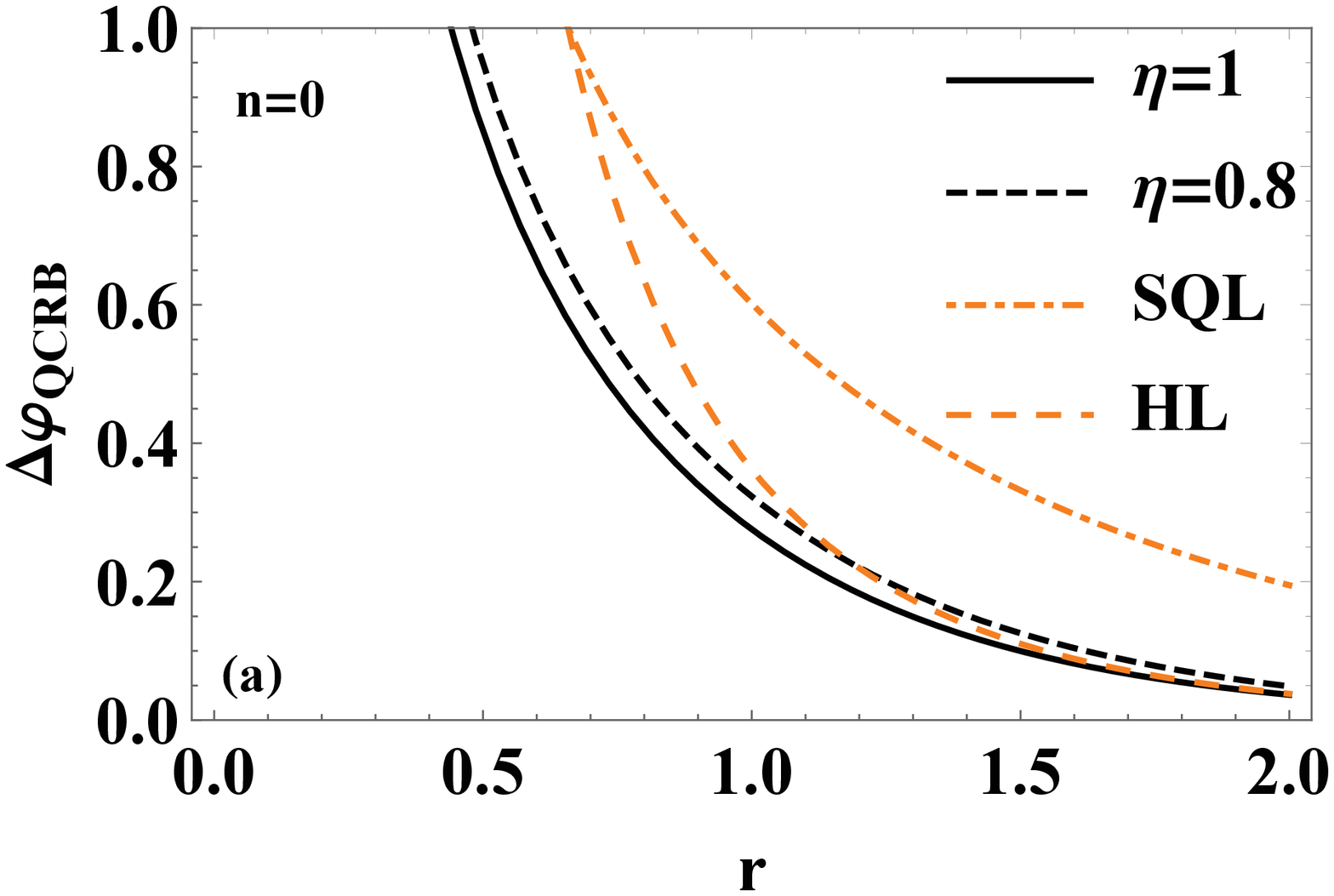}\includegraphics[width=0.5\textwidth]{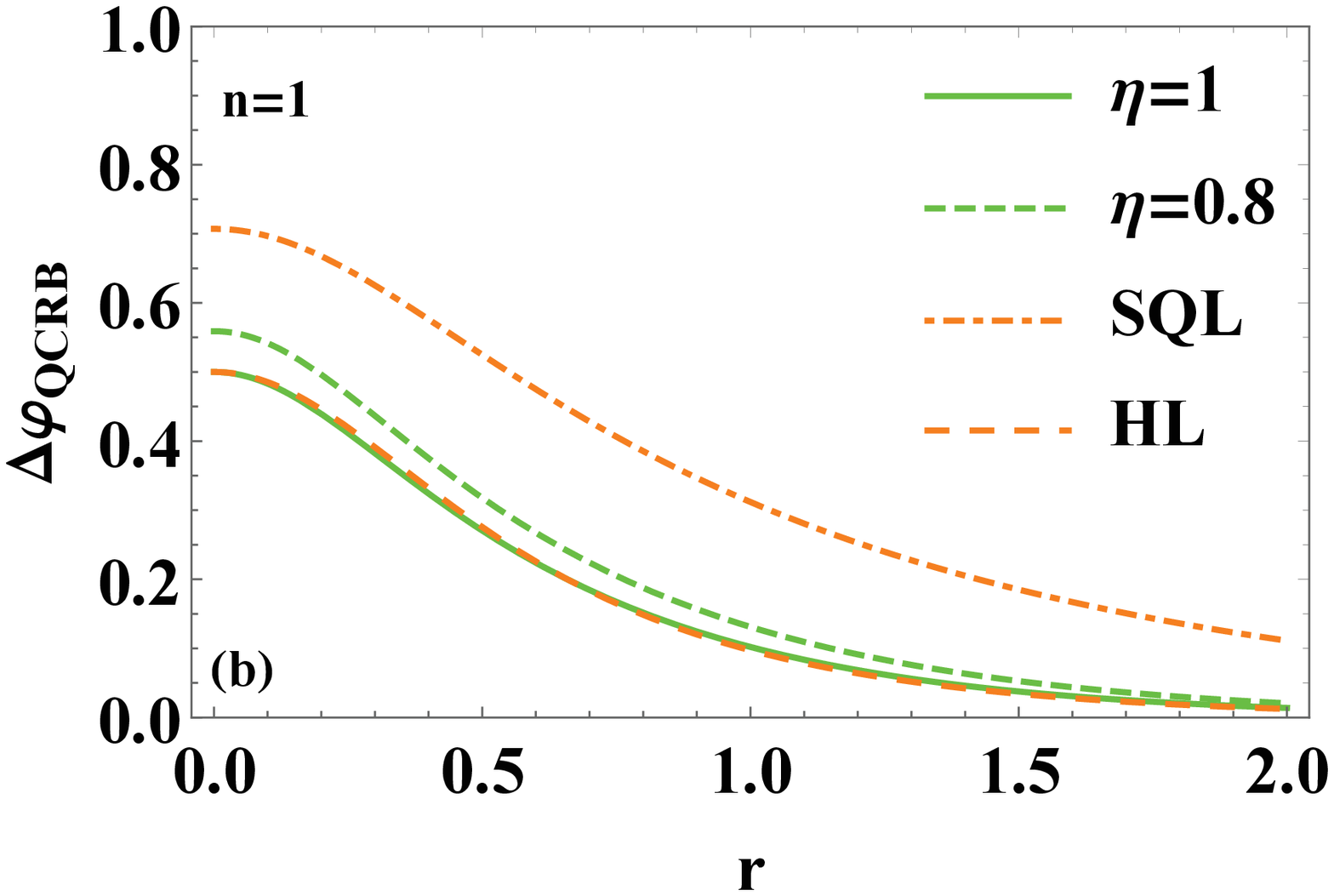}\\
\includegraphics[width=0.5\textwidth]{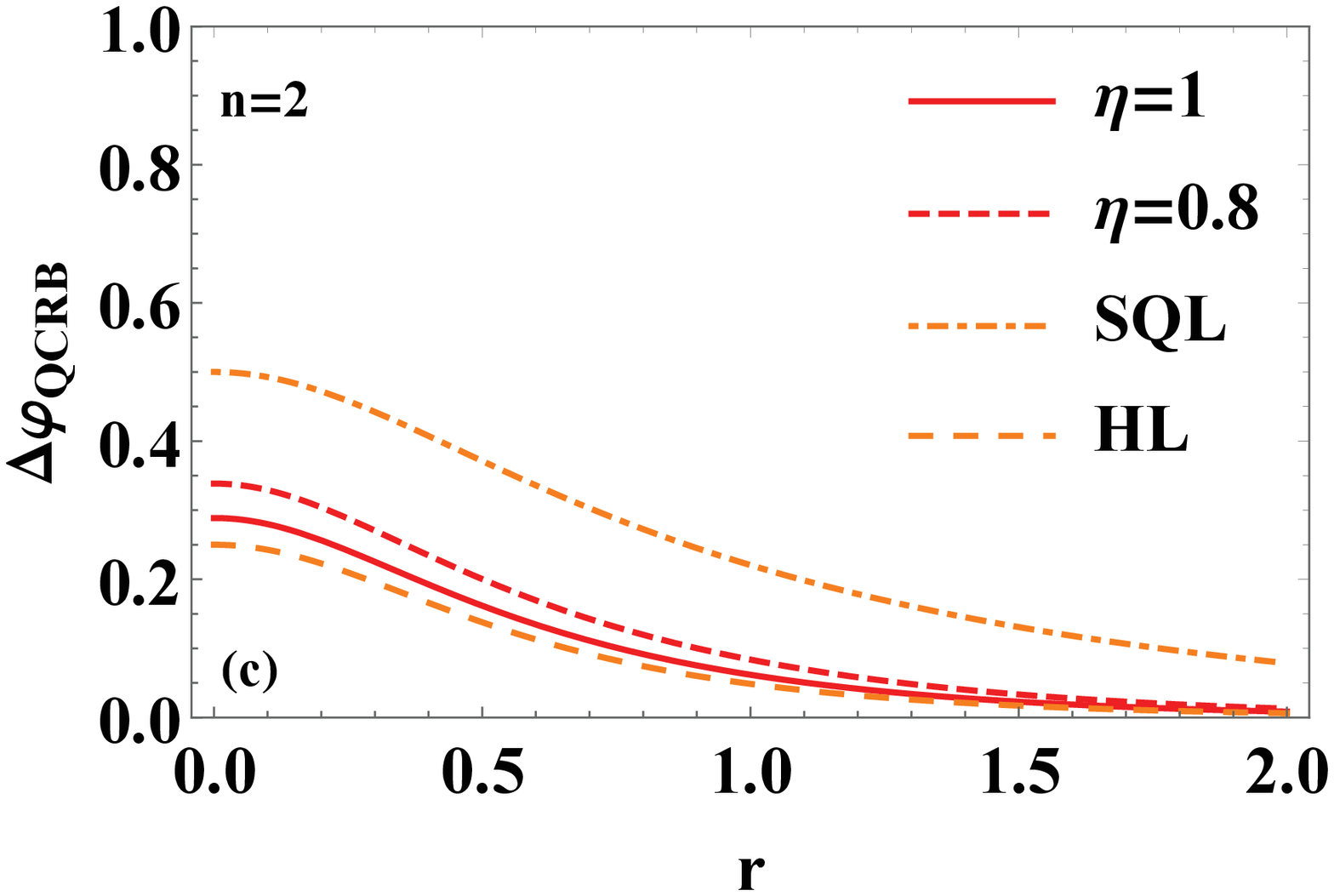}\includegraphics[width=0.5\textwidth]{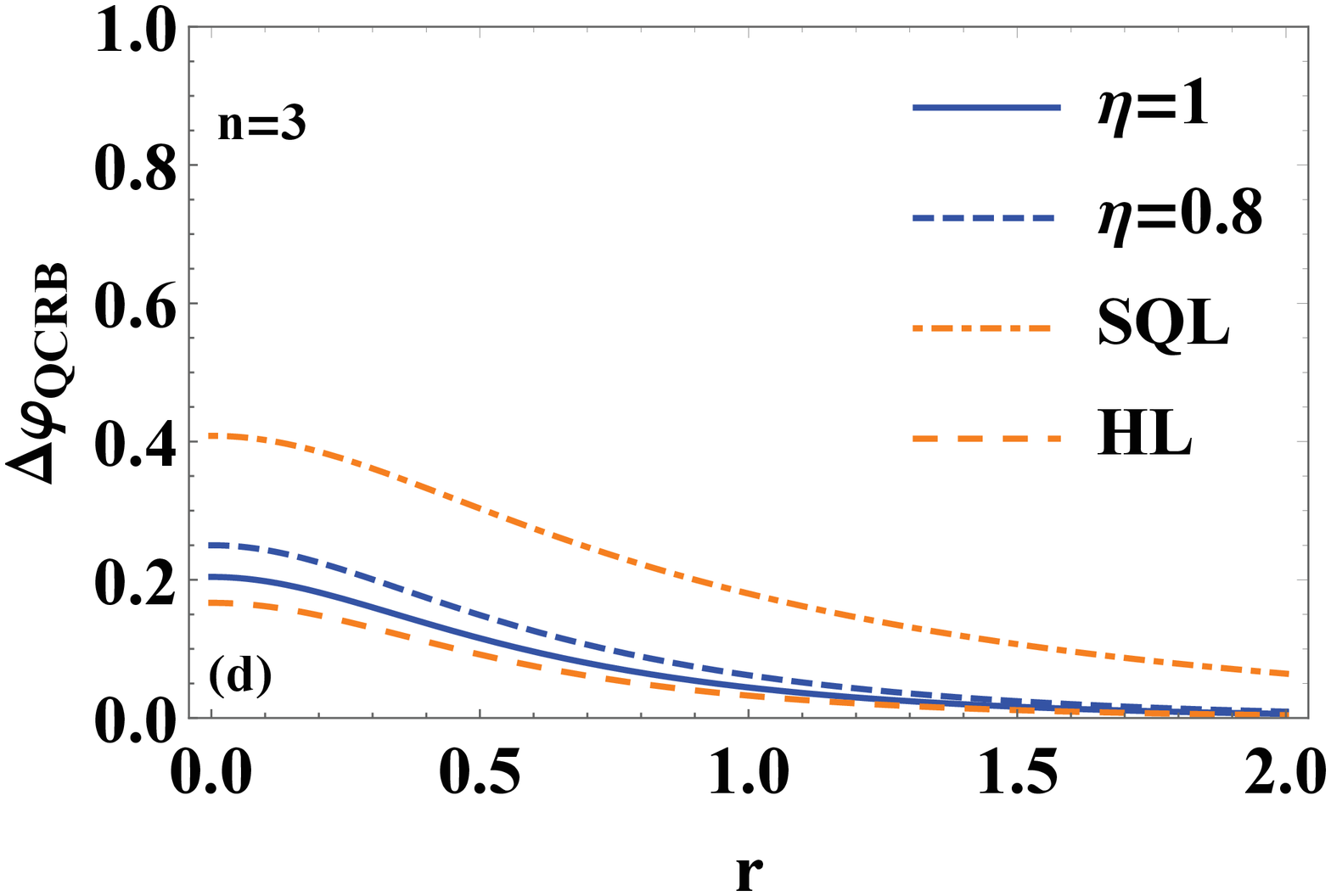}
\end{minipage}}
\caption{$\Delta \protect\varphi _{QCRB}$ as a function of the squeezing$\ $%
parameter$\ r$ comparing with the SQL and the HL (a) for $n=0$, $\protect%
\eta =1,0.8$, (b) for $n=1$, $\protect\eta =1,0.8$, (c)\ for $n=2$, $\protect%
\eta =1,0.8$, (d)\ for $n=3$, $\protect\eta =1,0.8$.}
\end{figure*}

Based on these formula above, we can clearly discuss the relation between
QFI and related parameters. In Fig. 10, we present the QFI $F_{Q}$ as a
function of the transmissivity $\eta $\ for different photon number $%
n=0,1,2,3$ under given the squeezing parameter ($r=0.7$). From Fig. 10, it
is clearly seen that the $F_{Q}$ increases with the increase of $\eta $ or $%
n $. This indicates that although photon losses can reduce the QFI, it can
be significantly improved by increasing the excited photon number. In
addition, with the increasing of $\eta $, the increasing of $n$ has more
clear improvement on the QFI.

Fig. 11 shows the QFI $F_{Q}$ and the QCRB as a function of the squeezing
parameter $r$ for the different transmissivity $\eta =1,0.8$\ and excited
photon number $n=0,1,2,3$. It is shown that the $F_{Q}$ can be increased by
increasing the excited photon number $n$\ or the squeezing parameter $r$. In
particular, the difference of $F_{Q}$ between ideal and photon-loss cases
increase as $n$\ or $r$, which becomes more clear in the small squeezing
region. This implies that, in a realistic case, the $F_{Q}$ with a higher
excited photon-number is more susceptible to the environment, especially in
small squeezing region. In addition, the case becomes less obvious in large
squeezing region. This case is true for the QCRB where $\Delta \varphi
_{QCRB}=1/\sqrt{F_{Q}}$, see Fig. 11(b).

In order to further clearly see the relation between the QCRB and the SQL,
the HL, we plot the QCRB $\Delta \varphi _{QCRB}$ as a function of the
squeezing parameter $r$ for different excited photon number $n=0,1,2,3$ and
both ideal and realistic cases in Fig. 12. Here the SQL and the HL are also
plotted for comparison. From Fig. 12, it is clear that (i) for the case of $%
n=0$ corresponding to the TMSV, the QCRB can break the SQL and the HL. In
fact, considering the TMSVs as inputs of the ideal MZI, it is found that $%
\Delta \varphi _{QCRB}=\frac{1}{\sqrt{\bar{N}^{2}+2\bar{N}}},$which exceeds
the HL defined as$\frac{1}{\bar{N}}$ \cite{27}. (ii) for the cases of $%
n=1,2,3$, the QCRB is between the SQL and the HL. In particular, for the
ideal case of $\eta =1$, the QCRB with $n=1$ basically coincides with the
HL. (iii) the QCRB can almost saturate the HL as the increasing of $r.$
Although the QCRB breaks the HL at $n=0$, the QCRB can still be improved
with increasing $n$. In addition, the difference between ideal and realistic
cases becomes smaller with increasing $r$. These results imply that although
the QCRB can break the HL for the TMSV, the QCRB can be further improved by
introducing excited photon number.

\section{Conclusion}

In summary, we introduced a kind of non-Gaussian state, i.e., Laguerre
polynomial excited squeezed state as input of the traditional MZI. Then we
first investigated the phase sensitivity with parity detection and the QFI
in ideal case. In particular, we derived an equivalent operator by using the
Weyl ordering invariance under similarity transformations, whose normal
ordering form is given. It is convenient to calculate the average of parity
operator using the equivalent operator for any input state of the
traditional MZI. This method is also effect when considering the realistic
case.

We further examined the effects of photon losses on the phase sensitivity,
including internal and external losses. It is found that the external loss
presents a bigger influence than the internal one. Moreover, the phase
sensitivity can be improved with the increase of the excited photon number $%
n $ for any squeezing parameter $r$. Specially speaking, the optimal phase
sensitivity is at the point with $\varphi =0$, and becomes better as $n$
increases for the ideal case. For the realistic case, however, the optimal
point of the phase sensitivity will deviate from $\varphi =0$, and the $%
\Delta \varphi $ value corresponding to optimal point decreases with the
increase of $n$. It is interesting that even in the realistic case, the
phase sensitivity still surpass that by the TMSV in the ideal case, but the
improved region of $\varphi $ becomes smaller with the increase of $n$. When
fixing the total input photon number, the phase sensitivity can also be
enhanced by increasing $n$ in the realistic case, although this case is not
true in the ideal case.

In addition, we investigated the effects of photon losses on the QFI. It is
shown that although the QFI will reduce due to the photon losses, it can
still increase as the squeezing parameter $r$ or the photon number $n$. But
the $F_{Q}$ with a higher excited photon-number is more susceptible to the
environment, which becomes more clear in the small squeezing region. The
QCRB $\Delta \varphi _{QCRB}$ can break the SQL and even beat the HL, which
can be further improved by introducing excited photon number. These results
can be effectively applied to improve the accuracy of phase measurement in
the realistic case.

\begin{acknowledgments}
This work is supported by the National Natural Science Foundation of China
(Grants No. 11964013 and No. 11664017) and the Training Program for Academic
and Technical Leaders of Major Disciplines in Jiangxi Province (No.
20204BCJL22053).
\end{acknowledgments}

\end{document}